\documentclass[aip,jcp,amsmath,reprint]{revtex4-1}
\usepackage{graphics}
\DeclareMathOperator{\Tr}{Tr}

\begin{document}

\title{Simple physics of the partly pinned fluid systems}

\author{Vincent Krakoviack}
\email{vincent.krakoviack@ens-lyon.fr}
\affiliation{Laboratoire de chimie, {\'E}cole normale sup{\'e}rieure
  de Lyon, 46 all{\'e}e d'Italie, 69364 Lyon cedex 07, France}

\date{\today}

\begin{abstract}
  In this paper, we consider some aspects of the physics of the partly
  pinned (PP) systems obtained by freezing in place particles in
  equilibrium bulk fluid configurations in the normal (nonglassy)
  state. We first discuss the configurational overlap and the
  disconnected density correlation functions, both in the homogeneous
  and heterogeneous cases, using the tools of the theory of adsorption
  in disordered porous solids. The relevant Ornstein-Zernike equations
  are derived, and asymptotic results valid in the regime where the
  perturbation due to the pinning process is small are
  obtained. Second, we consider the homogeneous PP lattice gas as a
  means to make contact between pinning processes in particle and spin
  systems and show that it can be straightforwardly mapped onto a
  random field Ising model with a strongly asymmetric bimodal
  distribution of the field. Possible implications of these results
  for studies of the glass transition based on PP systems are also
  discussed.
\end{abstract}


\maketitle

\section{Introduction}

A number of recent theoretical and computational studies in the field
of the glass transition have rested upon consideration of various
types of so-called partly pinned (PP) systems, a class of models
broadly characterized by their preparation protocol. Indeed, all PP
systems have in common that they are generated from a simple
statistical-mechanical model, such as a fluid or a spin assembly, in
equilibrium under given thermodynamic conditions, by freezing
(``pinning'') a subensemble of its constituents in individual states
(particle positions or spin values) corresponding to a randomly chosen
instantaneous configuration and letting the rest evolve under the
influence of the frozen objects, all interactions and thermodynamic
parameters being kept unchanged. Depending on the specific setup, such
PP systems might be homogeneous, if the frozen objects are evenly
distributed in the whole volume of the system, or heterogeneous, if
they are located in a predefined region of space, the free ones being
restricted to the complementary domain.

Because the systems from which they are prepared undergo thermal
fluctuations, the PP systems always display quenched randomness, due
to the frozen part. So, for any given starting statistical-mechanical
model, set of thermodynamic parameters, and specific freezing
procedure, a whole ensemble of PP systems is actually obtained. This
leads one to focus on quantities averaged over this ensemble, as is
standard in the study of quenched-disordered systems.
\cite{LifGrePasbook} Thanks to the peculiar preparation protocol of
the PP systems, the statistical properties of this ensemble are
straightforwardly related to those of the starting model, and
important simplifications follow. \cite{SchKobBin04JPCB,Kra10PRE,%
  CavGriVer10JSMTE,FraSem11book} For instance, the sample-averaged
configurational properties of a PP system exactly match those of the
system from which it derives. Also, the configuration of the free
components immediately after the pinning step is automatically a
representative equilibrium configuration of the corresponding PP
sample. These features have been put to good use with studies in
essentially three directions.

First, the PP fluid systems can be seen as models of confined fluids,
with the frozen particles forming the confining
environment. Accordingly, they have regularly appeared in
computational studies of the dynamics of fluids in confinement,
\cite{VirAraMed95PRL,VirMedAra95PRE,ChaJuaMed08PRE,SchKobBinPar02PMB,%
  SchKobBin02EL,SchKobBin03EPJE,SchKobBin04JPCB,Kim03EL,ChaJagYet04PRE,%
  MitErrTru06PRE,KimMiySai09EL,KimMiySai10EPJST,KimMiySai11JPCM,%
  FenMryPryFol09PRE,KarLerPro12PA,KobRolBer12NatPhys,KobRolBer12PhysProc,%
  KlaVog13JCP,KlaHenVog14JCP} where their peculiar properties open up
some interesting perspectives. Indeed, since the pinning process
leaves the average configurational properties of a PP fluid system
identical to those of the bulk fluid from which it is prepared,
meaning in particular a perfectly flat average density profile and
unperturbed pair correlations, one might argue that the study of PP
systems gives access to the intrinsic dynamical effects of
confinement, based on the fact that the possibility of interfering
secondary effects originating in confinement-induced structural
changes has been virtually eliminated by construction.
\cite{SchKobBinPar02PMB,SchKobBin02EL,SchKobBin03EPJE,SchKobBin04JPCB,%
  KlaVog13JCP} Moreover, the immediate availability of a
representative equilibrium configuration of the confined fluid just
after the pinning step implies that it is not necessary to equilibrate
the system after the confinement has been introduced. So, equilibrated
bulk configurations are all that is required to produce PP fluid
systems in equilibrium. This represents a decisive advantage when one
is interested in the effect of confinement on the dynamics of
glassforming liquids, which are already slow and challenging to
investigate in the bulk and become much slower, sometimes by orders of
magnitude, after pinning, so that they would be completely out of
reach if equilibration was required. Thanks to this property, this
pinning strategy has turned out particularly fruitful for the study of
spatial dynamical correlations in confined glassy systems.
\cite{SchKobBinPar02PMB,SchKobBin02EL,SchKobBin03EPJE,%
  SchKobBin04JPCB,Kim03EL,KarLerPro12PA,KobRolBer12NatPhys,%
  KobRolBer12PhysProc}

A second point of view is to consider the pinning process as a
constraint imposed on the original statistical-mechanical model. It
has been argued that such a constraint, whose characteristics are by
construction fixed by the statistical properties of the system under
study itself, represents a means to probe the existence of nontrivial
static correlations in this system, without having to know their
precise microscopic nature \emph{a priori}. The basic idea is then to
measure how the frozen degrees of freedom influence the accessible
states of the free ones in the PP system, via the investigation of
point-to-set correlations. \cite{BouBir04JCP,MonSem06JSP_2,%
  MezMon06JSP,MonSem06JSP,FraMon07JPA,ZarFra10JSM,FraSem11book} This
insight has motivated numerous computational studies of glassy systems
in a variety of pinning geometries. \cite{JacGar05JCP,CavGriVer07PRL,%
  BirBouCavGriVer08NatPhys,CavGriVer10JSMTE,CavGriVer12JCP,%
  GraTroCavGriVer13JCP,SauLev11PRL,HocMarRei12PRL,BerKob12PRE,%
  KobRolBer12NatPhys,KobRolBer12PhysProc,ChaChaTar12PRL,ChaChaTar13JCP,%
  ChaTar13PRE,BirKarPro13PRL,LiXuSun14JCP}

Finally, the PP systems can be considered as mere extensions of the
statistical-mechanical model from which they derive, with an enlarged
parameter space thanks to the addition as state variables of
quantitative descriptors of the pinning process, such as the fraction
of pinned constituents in the homogeneous case. Then, one can explore
these extra dimensions with various models and theoretical tools,
trying to unveil novel physical phenomena and possibly contrasting
scenarios that would allow one to compare different approaches. Such
studies have been recently undertaken for glassy systems, in the
framework of the random first-order transition (RFOT) theory,
\cite{CamBir12EPL,CamBir12PNAS,CamBir13JCP,FraParRic13JSMTE,CamGraBir13PRL}
of the mode-coupling theory (MCT), \cite{Kra11PRE,SzaFle13EPL} and of
some dynamical facilitation models, \cite{JacBer12PRE} and in computer
simulations of fluid models. \cite{KimMiySai09EL,%
  KimMiySai10EPJST,KimMiySai11JPCM,KobBer13PRL,JacFul13PRE}

From this overview, it is manifest that a lot has been done to
characterize the complex physics of the PP systems in the glassy
regime. In comparison, their nonglassy physics is far less
documented. This is precisely the focus of the present work, in which
aspects of the theory of the PP systems deriving from normal fluid
states are discussed. The motivation behind this study is that these
systems represent a natural reference with respect to which the
behavior of their glassy counterparts can be assessed. Thus, a better
knowledge of their properties offers the prospect of pinpointing and
maybe disentangling situations in which features that are mere
consequences of the partial pinning approach possibly coexist and
interfere with others originating in glassiness.

The paper is organized as follows. In Sec.~\ref{sec:basic}, basic
properties of the PP systems needed in the following are reported. The
main results of Ref.~\onlinecite{Kra10PRE} on the homogeneous case are
summarized, then extended to deal with heterogeneous pinning schemes.
In Sec.~\ref{sec:overlap}, the configurational overlap function
introduced in the study of constrained systems is considered from the
point of view of the theory of adsorption in disordered porous solids,
in which many powerful tools have been developed to deal with simple
fluids in contact with rigid quenched-disordered substrates.
\cite{MadGla88JSP,Mad92JCP,GivSte92JCP,LomGivSteWeiLev93PRE,%
  GivSte94PA,RosTarSte94JCP,DonKieRos94PRE,Sch02PRE,Sch03PRE,ReiSch04JSP,%
  Sch09PRE,LafCue06PRE} The homogeneous PP lattice gas is introduced
in Sec.~\ref{sec:lattice} as a means to make contact between pinning
processes in particle and spin systems. In both of the latter
sections, possible implications of the results for studies of glassy
PP systems are discussed. Finally, Sec.~\ref{sec:conclusion} is
devoted to concluding remarks.
\section{Basic properties of the partly pinned fluid systems}
\label{sec:basic}

In this Section are reviewed simple properties of the PP fluid systems
that are useful in the following.

As described in the Introduction, the PP fluid systems are generated
from equilibrium bulk fluid configurations by pinning in place part of
their particles, so that a quenched-disordered solid substrate is
formed under whose influence evolve the remaining particles. We shall
refer to the mobile and immobile particles as confined fluid (index f)
and matrix (index m), respectively.

Although this is not strictly necessary, it is convenient to introduce
a third species called template (index t), by reference to the
templated-depleted systems studied by Van Tassel \textit{et al}.
\cite{TasTalVioTar97PRE,Tas97JCP,Tas99PRE,ZhaTas00JCP,ZhaTas00MP,%
  ZhaCheTas01PRE} Prior to pinning, the template refers to the
particles of the original bulk system that will be left unpinned and
can be seen as a precursor of the confined fluid. After pinning, it
consists of immobile ``ghost'' particles marking the positions of
these unpinned particles at the time of pinning. The template is
essentially useful as a bookkeeping device in order to measure how the
confined fluid configurations remain correlated with the bulk
configurations from which the PP systems are prepared.

For simplicity, a grand-canonical description is used. Thus, the
starting point is a one-component bulk fluid in a volume $V$ at
temperature $T$ and activity $z$. \cite{macdohansen3ed} As usual, we
define $\beta=1/k_\text{B}T$, $k_\text{B}$ being Boltzmann's
constant. The probability density that this system consists of $N$
particles located at $(\mathbf{x}_1, \mathbf{x}_2, \ldots,
\mathbf{x}_N) \equiv \mathbf{x}^{N}$ is
\begin{equation}\label{pbulk}
  \mathcal{P}_\text{bulk}(N,\mathbf{x}^{N})=%
  \frac{z^{N}e^{-\beta V_\text{bulk}(N,\mathbf{x}^{N})}}%
  {\Xi_\text{bulk}N!},
\end{equation}
where $V_\text{bulk}(N,\mathbf{x}^{N})$ denotes the potential
energy. $\Xi_\text{bulk}$ is the partition function, which, with the
shorthand notation $\Tr\cdots\equiv\sum_{N=0}^{+\infty} \int
d\mathbf{x}^N\cdots$, reads
\begin{equation}
  \Xi_\text{bulk}=\Tr\frac{z^{N}e^{-\beta%
      V_\text{bulk}(N,\mathbf{x}^{N})}}{N!}.
\end{equation} 

\subsection{Homogeneous systems}
\label{subsec:reminder}

The ensemble of homogeneous PP fluid systems with average pinning
fraction $x$ is generated by imposing that, for every configuration of
the fluid, each particle in the system can be frozen in place or not,
with probabilities $x$ and $1-x$, respectively.  The resulting joint
probability density that one produces a matrix with $N_\text{m}$
pinned particles located at $(\mathbf{q}_1, \mathbf{q}_2, \ldots,
\mathbf{q}_{N_\text{m}}) \equiv \mathbf{q}^{N_\text{m}}$ while
$N_\text{t}$ unpinned particles are located at $(\mathbf{s}_1,
\mathbf{s}_2, \ldots, \mathbf{s}_{N_\text{t}}) \equiv
\mathbf{s}^{N_\text{t}}$, is thus
\begin{multline}\label{pinjoint}
  \mathcal{P}_\text{mt}(N_\text{m},\mathbf{q}^{N_\text{m}},
  N_\text{t},\mathbf{s}^{N_\text{t}}) = \\
  \frac{x^{N_\text{m}} (1-x)^{N_\text{t}} z^{N_\text{m}+N_\text{t}}
    e^{ -\beta V_\text{bulk}(N_\text{m}+N_\text{t},
      \mathbf{q}^{N_\text{m}},\mathbf{s}^{N_\text{t}})}}%
  {\Xi_\text{bulk} N_\text{m}!N_\text{t}!}.
\end{multline}
It follows that the matrix configurations are distributed according to
the probability density
\begin{equation} \label{pinmatrix}
  \mathcal{P}_\text{m}(N_\text{m},\mathbf{q}^{N_\text{m}}) =
  \Tr_\text{t} \mathcal{P}_\text{mt}(N_\text{m},\mathbf{q}^{N_\text{m}},
  N_\text{t},\mathbf{s}^{N_\text{t}}). 
\end{equation}

In order to now consider the confined fluid, it is necessary to
recognize that Eq.~\eqref{pinjoint} has the form suitable for a bulk
ideal binary mixture in which the matrix and the template have
activities $xz$ and $(1-x)z$, respectively, and to separate the
matrix-matrix, matrix-template, and template-template contributions to
the potential energy prior to pinning, so that
\begin{multline}\label{eq:potsplit}
  V_\text{bulk}(N_\text{m}+N_\text{t},\mathbf{q}^{N_\text{m}},%
  \mathbf{s}^{N_\text{t}})=V_\text{mm}(N_\text{m},\mathbf{q}^{N_\text{m}})
  \\+V_\text{mt}(N_\text{m},\mathbf{q}^{N_\text{m}},%
  N_\text{t},\mathbf{s}^{N_\text{t}})+%
  V_\text{tt}(N_\text{t},\mathbf{s}^{N_\text{t}}).
\end{multline}
Then, since the confined fluid is assumed by construction to inherit
its characteristics from the unpinned particles, its activity is taken
equal to $(1-x)z$ and its potential energy to
$V_\text{mt}(N_\text{m},\mathbf{q}^{N_\text{m}},N_\text{f},
\mathbf{r}^{N_\text{f}})+V_\text{tt}(N_\text{f},\mathbf{r}^{N_\text{f}})$,
when $N_\text{f}$ fluid particles located at $(\mathbf{r}_1,
\mathbf{r}_2, \ldots, \mathbf{r}_{N_\text{f}}) \equiv
\mathbf{r}^{N_\text{f}}$ are in presence of a matrix with $N_\text{m}$
particles located at $\mathbf{q}^{N_\text{m}}$. The probability
density of such a fluid configuration depends parametrically on the
matrix realization and reads
\begin{multline} \label{fluidprob}
  \mathcal{P}_\text{f}(N_\text{f},\mathbf{r}^{N_\text{f}} |
  N_\text{m},\mathbf{q}^{N_\text{m}}) = \\
  \frac{[(1-x)z]^{N_\text{f}} e^{ -\beta \left[
        V_\text{mt}(N_\text{m}, \mathbf{q}^{N_\text{m}}, N_\text{f},
        \mathbf{r}^{N_\text{f}}) + V_\text{tt}(N_\text{f},
        \mathbf{r}^{N_\text{f}}) \right]
    }}{\Xi_\text{f}(N_\text{m},\mathbf{q}^{N_\text{m}}) N_\text{f}!},
\end{multline}
with the confined fluid partition function
\begin{multline}
  \Xi_\text{f}(N_\text{m},\mathbf{q}^{N_\text{m}}) = \\ \Tr_\text{f}
  \frac{[(1-x)z]^{N_\text{f}} e^{ -\beta \left[
        V_\text{mt}(N_\text{m}, \mathbf{q}^{N_\text{m}}, N_\text{f},
        \mathbf{r}^{N_\text{f}}) + V_\text{tt}(N_\text{f},
        \mathbf{r}^{N_\text{f}}) \right]}}{N_\text{f}!}.
\end{multline}

These probability distributions obey the identity
\begin{multline} \label{identity} %
  \mathcal{P}_\text{mt}(N_\text{m}, \mathbf{q}^{N_\text{m}},
  N_\text{t}, \mathbf{s}^{N_\text{t}})
  \mathcal{P}_\text{f}(N_\text{f}, \mathbf{r}^{N_\text{f}} |
  N_\text{m}, \mathbf{q}^{N_\text{m}}) = \\
  \mathcal{P}_\text{mt}(N_\text{m}, \mathbf{q}^{N_\text{m}},
  N_\text{f},\mathbf{r}^{N_\text{f}} )
  \mathcal{P}_\text{f}(N_\text{t},\mathbf{s}^{N_\text{t}} |
  N_\text{m},\mathbf{q}^{N_\text{m}}),
\end{multline}
from which the characteristic configurational properties of the PP
systems follow. \cite{SchKobBin04JPCB,Kra10PRE,CavGriVer10JSMTE,%
  FraSem11book} Thus, if $\rho$ and $h(r)$ are the number density and
total correlation function of the original bulk fluid, one gets for
the corresponding quantities pertaining to the different species in
the PP system, \footnote{These quantities are defined in terms of the
  usual one- and two-body density operators of the theory of mixtures,
  with configurational averages taken over the composite probability
  distribution $\mathcal{P}_\text{mt}(N_\text{m},
  \mathbf{q}^{N_\text{m}}, N_\text{t}, \mathbf{s}^{N_\text{t}})
  \mathcal{P}_\text{f}(N_\text{f}, \mathbf{r}^{N_\text{f}} |
  N_\text{m}, \mathbf{q}^{N_\text{m}})$. See
  Ref.~\onlinecite{Kra10PRE} for details.}
\begin{subequations} \label{eq:identhomo}
  \begin{gather}
    \rho_\text{m} = x \rho, \\
    \rho_\text{t} = \rho_\text{f} = (1-x) \rho, \\
    h_\text{mm}(r) = h_\text{mt}(r) = h_\text{mf}(r) = h_\text{tt}(r)
    = h_\text{f\/f}(r) =  h(r), \\
    h_\text{tf}(r) = h_\text{dis}(r).
  \end{gather}
\end{subequations}
In the last equation, $h_\text{dis}(r)$ denotes the disconnected total
correlation function that will be defined and discussed in detail in
Sec.~\ref{sec:overlap}.

\subsection{Heterogeneous systems}

In a heterogeneous PP fluid system, the volume occupied by the
original bulk fluid is partitioned into two predefined complementary
regions denoted by $\mathcal{R}$ and $\overline{\mathcal{R}}$. In
principle, they can be of arbitrary shape, but simple geometries are
obviously favored in practical calculations. \cite{SchKobBinPar02PMB,%
  SchKobBin02EL,SchKobBin03EPJE,SchKobBin04JPCB,CavGriVer07PRL,%
  BirBouCavGriVer08NatPhys,CavGriVer10JSMTE,CavGriVer12JCP,%
  GraTroCavGriVer13JCP,KobRolBer12NatPhys,KobRolBer12PhysProc,%
  SauLev11PRL,HocMarRei12PRL,BerKob12PRE,BirKarPro13PRL,KlaVog13JCP,%
  KlaHenVog14JCP,LiXuSun14JCP} Then, the corresponding ensemble of PP
systems is generated by imposing that, for every configuration of the
fluid, the particles located in $\overline{\mathcal{R}}$ are pinned,
while those in $\mathcal{R}$ are left mobile.

A convenient mathematical tool encoding the region $\mathcal{R}$ is
its indicator or characteristic function, defined as
\begin{equation}
  \chi_\mathcal{R}(\mathbf{x}) = 
  \begin{cases}
    1 & \text{ if $\mathbf{x}\in\mathcal{R}$,} \\
    0 & \text{ if $\mathbf{x}\not\in\mathcal{R}$.}
  \end{cases}
\end{equation}
The corresponding function for $\overline{\mathcal{R}}$ is obviously
$\chi_{\overline{\mathcal{R}}}(\mathbf{x}) = 1 -
\chi_\mathcal{R}(\mathbf{x})$. \footnote{The introduction of these
  indicator functions is an alternative to the splitting of the
  configurational integrals used by Scheidler \emph{et al.} in
  Ref.~\onlinecite{SchKobBin04JPCB}.} Then, starting from
Eq.~\eqref{pbulk} and according to the description of the
heterogeneous pinning process, the joint probability density that one
obtains a random substrate with $N_\text{m}$ pinned particles located
at $\mathbf{q}^{N_\text{m}}$ while $N_\text{t}$ unpinned particles are
located at $\mathbf{s}^{N_\text{t}}$ reads
\begin{multline}\label{pinjointhetero}
  \mathcal{P}_\text{mt}(N_\text{m},\mathbf{q}^{N_\text{m}},
  N_\text{t},\mathbf{s}^{N_\text{t}}) =
  \frac{z^{N_\text{m}+N_\text{t}}}{\Xi_\text{bulk}
    N_\text{m}!N_\text{t}!} \\
  \times e^{-\beta V_\text{bulk}(N_\text{m}+N_\text{t},
    \mathbf{q}^{N_\text{m}},\mathbf{s}^{N_\text{t}})}
  \prod_{i=1}^{N_\text{m}} \chi_{\overline{\mathcal{R}}}(\mathbf{q}_i)
  \prod_{j=1}^{N_\text{t}} \chi_\mathcal{R}(\mathbf{s}_j).
\end{multline}
Clearly, this probability density is identically zero if particles in
$\overline{\mathcal{R}}$ are left unpinned and/or particles in
$\mathcal{R}$ are pinned. It therefore properly describes the targeted
realization ensemble. Here as well, the probability distribution
$\mathcal{P}_\text{m}(N_\text{m},\mathbf{q}^{N_\text{m}})$ of the
substrate configurations follows from Eq.~\eqref{pinmatrix}.

The normalization condition of
$\mathcal{P}_\text{mt}(N_\text{m},\mathbf{q}^{N_\text{m}},
N_\text{t},\mathbf{s}^{N_\text{t}})$ allows one to get
\begin{multline}
  \Xi_\text{bulk} = \Tr_\text{m} \Tr_\text{t}
  \frac{z^{N_\text{m}+N_\text{t}}}{N_\text{m}!N_\text{t}!} \\
  \times e^{ -\beta V_\text{bulk}(N_\text{m}+N_\text{t},
    \mathbf{q}^{N_\text{m}},\mathbf{s}^{N_\text{t}})}
  \prod_{i=1}^{N_\text{m}} \chi_{\overline{\mathcal{R}}}(\mathbf{q}_i)
  \prod_{j=1}^{N_\text{t}} \chi_\mathcal{R}(\mathbf{s}_j).
\end{multline}
This corresponds to a rather unusual reformulation of the bulk system
in terms of two subsystems, similar in spirit to the reformulation in
terms of an ideal binary mixture in the homogeneous
case. \cite{Kra10PRE} Here, since the indicator functions can be
interpreted as Boltzmann factors corresponding to exclusion
potentials, the two subsystems are spatially disjoint and the above
expression of $\Xi_\text{bulk}$ features the coexistence of a matrix
component confined to $\overline{\mathcal{R}}$ with a template
component confined to $\mathcal{R}$, both with activity $z$. A similar
construction, but in the canonical ensemble, can be found in
Ref.~\onlinecite{CavGriVer10JSMTE}.

Once again, the confined fluid component in the PP system should
inherit its characteristics from the template. So, with the same
splitting of the potential energy as in the homogeneous case,
Eq.~\eqref{eq:potsplit}, the probability density of a fluid
configuration with $N_\text{f}$ particles at $\mathbf{r}^{N_\text{f}}$
in presence of a substrate with $N_\text{m}$ particles at
$\mathbf{q}^{N_\text{m}}$ reads
\begin{multline}
  \mathcal{P}_\text{f}(N_\text{f},\mathbf{r}^{N_\text{f}} |
  N_\text{m},\mathbf{q}^{N_\text{m}}) =
  \frac{z^{N_\text{f}}}{\Xi_\text{f}(N_\text{m},\mathbf{q}^{N_\text{m}})
    N_\text{f}!} \\
  \times e^{ -\beta \left[ V_\text{mt}(N_\text{m},
      \mathbf{q}^{N_\text{m}}, N_\text{f}, \mathbf{r}^{N_\text{f}}) +
      V_\text{tt}(N_\text{f}, \mathbf{r}^{N_\text{f}}) \right]}
  \prod_{i=1}^{N_\text{f}} \chi_\mathcal{R}(\mathbf{r}_i),
\end{multline}
with the confined fluid partition function
\begin{multline}
  \Xi_\text{f}(N_\text{m},\mathbf{q}^{N_\text{m}}) = \Tr_\text{f}
  \frac{z^{N_\text{f}}}{N_\text{f}!} \\
  \times e^{ -\beta \left[ V_\text{mt}(N_\text{m},
      \mathbf{q}^{N_\text{m}}, N_\text{f}, \mathbf{r}^{N_\text{f}}) +
      V_\text{tt}(N_\text{f}, \mathbf{r}^{N_\text{f}}) \right]}
  \prod_{i=1}^{N_\text{f}} \chi_\mathcal{R}(\mathbf{r}_i).
\end{multline}

It is straightforward to check that these probability distributions
obey Eq.~\eqref{identity}, so that the configurational identities of
Ref.~\onlinecite{Kra10PRE} (some of them have been first reported in
Ref.~\onlinecite{SchKobBin04JPCB}) and their consequences can be
readily extended to deal with the present type of systems. In
particular, one gets for the density profiles and total correlation
functions of the different species, \cite{Note1}
\begin{subequations} \label{eq:identhetero}
  \begin{gather}
    \rho_\text{m}(\mathbf{x}) =
    \chi_{\overline{\mathcal{R}}}(\mathbf{x}) \rho, \\
    \rho_\text{t}(\mathbf{x}) = \rho_\text{f}(\mathbf{x}) =
    \chi_{\mathcal{R}}(\mathbf{x}) \rho, \label{eq:profile} \\
    h_\text{mm}(\mathbf{x},\mathbf{y}) = h(|\mathbf{x}-\mathbf{y}|),
    \quad \mathbf{x},\mathbf{y} \in \overline{\mathcal{R}}, \\
    h_\text{mt}(\mathbf{x},\mathbf{y}) =
    h_\text{mf}(\mathbf{x},\mathbf{y}) = h(|\mathbf{x}-\mathbf{y}|),
    \quad \mathbf{x} \in \overline{\mathcal{R}}, \mathbf{y} \in
    \mathcal{R}, \\
    h_\text{tt}(\mathbf{x},\mathbf{y}) =
    h_\text{ff}(\mathbf{x},\mathbf{y}) = h(|\mathbf{x}-\mathbf{y}|),
    \quad \mathbf{x},\mathbf{y} \in \mathcal{R}, \\
    h_\text{tf}(\mathbf{x},\mathbf{y}) =
    h_\text{dis}(\mathbf{x},\mathbf{y}), \quad \mathbf{x},\mathbf{y}
    \in \mathcal{R},
  \end{gather}
\end{subequations}
when the pinning process is performed on a bulk fluid with density
$\rho$ and total correlation function $h(r)$. Again, in the last
equation, $h_\text{dis}(\mathbf{x},\mathbf{y})$ denotes the
disconnected total correlation function that will be studied in
Sec.~\ref{sec:overlap}.

We close this section with two remarks of practical use.

The present derivation clearly shows that the potential energy of the
confined fluid in a heterogeneous PP system must contain an infinite
contribution excluding the fluid particles from the matrix
domain. Otherwise, without the corresponding product of
$\chi_\mathcal{R}$ functions in
$\mathcal{P}_\text{f}(N_\text{f},\mathbf{r}^{N_\text{f}} |
N_\text{m},\mathbf{q}^{N_\text{m}})$, Eq.~\eqref{identity} is not
obeyed and the ensuing characteristic configurational properties of
the PP systems \cite{SchKobBin04JPCB,Kra10PRE,CavGriVer10JSMTE,%
  FraSem11book} are lost. Such a contribution has been included in
most computer simulation studies of heterogeneous PP systems on
empirical grounds, in order to prevent the invasion of the substrate
by the fluid particles. We find here that it is actually an integral
part of the proper definition of a heterogeneous PP system.

The equality $h_\text{ff}(\mathbf{x},\mathbf{y}) =
h(|\mathbf{x}-\mathbf{y}|)$, $\mathbf{x},\mathbf{y} \in \mathcal{R}$,
is also worth a comment. Indeed, it is valid even if the points
$\mathbf{x}$ and $\mathbf{y}$ belong to disconnected domains of
$\mathcal{R}$, for instance, if they are separated from each other by
a slab of pinned particles. It therefore implies that, in order to
ensure the statistical independence of the two interfacial subsystems
located on each side of such a separation, a minimal requirement is in
principle that this separation be thick enough to allow a complete
decay of $h(|\mathbf{x}-\mathbf{y}|)$ between its boundaries. For
dense fluids, this generically means thicknesses significantly larger
than the mere range of the interaction potential.
\section{Overlap and disconnected correlation functions in the
  nonglassy partly pinned fluid systems}
\label{sec:overlap}

In this Section, we consider the PP fluid systems from the point of
view of the theory of simple fluids adsorbed on disordered substrates.
\cite{MadGla88JSP,Mad92JCP,GivSte92JCP,LomGivSteWeiLev93PRE,%
  GivSte94PA,RosTarSte94JCP,DonKieRos94PRE,Sch02PRE,Sch03PRE,%
  ReiSch04JSP,Sch09PRE,LafCue06PRE} We first show that the
configurational overlap function is simply related to the disconnected
two-point density and total correlation function. We then report some
potentially interesting properties of the latter. Finally, we discuss
how this knowledge could be useful in studies of glassy PP systems.

\subsection{Measures of the one-body fluid density}

As pointed out in the Introduction, the study of the PP fluid systems
involves a disorder average over the matrix realizations in addition
to the ordinary Boltzmann-Gibbs average which is then
realization-dependent. It follows that specific types of
configurational quantities, characteristic of random systems, come
into play, as is already visible when dealing with the one-body fluid
density, the most fundamental descriptor of the microscopic structure
of a fluid in an inhomogeneous external
potential. \cite{macdohansen3ed}

In any single sample of a PP fluid system, the randomly placed matrix
particles are the sources of a very complex potential energy landscape
in which the fluid is plunged. A fluid density profile results,
\begin{equation}
  \rho_\text{f}[N_\text{m},\mathbf{q}^{N_\text{m}}](\mathbf{x})
  = \left\langle \hat{\rho}_\text{f}(\mathbf{x};
    N_\text{f},\mathbf{r}^{N_\text{f}})
  \right\rangle_{N_\text{m},\mathbf{q}^{N_\text{m}}},
\end{equation}
corresponding to the realization-dependent thermal average, denoted by
$\langle \cdots \rangle_{N_\text{m},\mathbf{q}^{N_\text{m}}}$, of the
microscopic one-body fluid density operator
\begin{equation}
  \hat{\rho}_\text{f}(\mathbf{x};
  N_\text{f},\mathbf{r}^{N_\text{f}}) = \sum_{i=1}^{N_\text{f}}
  \delta(\mathbf{x} - \mathbf{r}_i).
\end{equation}
Because of its parametric dependence on $N_\text{m}$ and
$\mathbf{q}^{N_\text{m}}$, $\rho_\text{f}[N_\text{m},
\mathbf{q}^{N_\text{m}}](\mathbf{x})$ is a highly complicated function
of $\mathbf{x}$, with significant variations at the scale of the
particle size.

From the point of view of the realization ensemble,
$\rho_\text{f}[N_\text{m},\mathbf{q}^{N_\text{m}}] (\mathbf{x})$
appears as a scalar random field. A basic characterization of such an
object is then through its mean (already considered in the previous
section)
\begin{equation}
  \rho_\text{f}(\mathbf{x}) = \overline{\rho_\text{f}
    [N_\text{m},\mathbf{q}^{N_\text{m}}](\mathbf{x})}
\end{equation}
and correlation function, often called the disconnected two-point
density,
\begin{equation}\label{eq:discdens}
  \rho_\text{dis}(\mathbf{x},\mathbf{y}) =
  \overline{\rho_\text{f}[N_\text{m},\mathbf{q}^{N_\text{m}}]
    (\mathbf{x})
    \rho_\text{f}[N_\text{m},\mathbf{q}^{N_\text{m}}]
    (\mathbf{y})},
\end{equation}
where $\overline{\cdots}$ denotes the disorder average. Through this
averaging, these functions usually acquire a simple spatial dependence
reflecting the symmetries of the pinning process. For instance, in the
case of a homogeneous pinning, the average density profile
$\rho_\text{f}(\mathbf{x})$ is actually independent of $\mathbf{x}$
and $\rho_\text{dis}(\mathbf{x},\mathbf{y})$ is a function of
$|\mathbf{x}-\mathbf{y}|$ only.

Coarse-grained density fields might also be considered. Then, space is
decomposed into cells, and the coarse-grained fluid density operator
in cell $i$ located at $\mathbf{R}_i$ is defined as
\begin{equation}
  \hat{n}_i(N_\text{f},\mathbf{r}^{N_\text{f}}) =
  \int_\text{cell $i$} \!\!\!\!\!\!\! d\mathbf{u}_i
  \hat{\rho}_\text{f}(\mathbf{R}_i+\mathbf{u}_i;
  N_\text{f},\mathbf{r}^{N_\text{f}}),
\end{equation}
from which the analogues of the above functions are built,
\begin{gather}
  n_i[N_\text{m},\mathbf{q}^{N_\text{m}}] = \left\langle
    \hat{n}_i(N_\text{f},\mathbf{r}^{N_\text{f}})
  \right\rangle_{N_\text{m},\mathbf{q}^{N_\text{m}}}, \\
  n_i = \overline{n_i[N_\text{m},\mathbf{q}^{N_\text{m}}]}, \\
  n_{ij,\text{dis}} = \overline{
    n_i[N_\text{m},\mathbf{q}^{N_\text{m}}]
    n_j[N_\text{m},\mathbf{q}^{N_\text{m}}]}.
\end{gather}

It is immediate that
\begin{equation} \label{eq:overlapfromdis} %
  n_{ij,\text{dis}} = \int_\text{cell $i$} \!\!\!\!\!\!\!
  d\mathbf{u}_i \int_\text{cell $j$ } \!\!\!\!\!\!\!  d\mathbf{u}_j
  \rho_\text{dis}(\mathbf{R}_i+\mathbf{u}_i,
  \mathbf{R}_j+\mathbf{u}_j),
\end{equation}
and this equation is actually the starting point of a generic
grid-based method that can be used to evaluate the disconnected
two-point density in computer simulations \cite{MerLevWei96JCP} (see
Ref.~\onlinecite{SchCosKurKah09MP} for a similar analysis in
reciprocal space). One can also check that $\hat{n}_i(N_\text{f},
\mathbf{r}^{N_\text{f}})$ is equal to the number of fluid particles in
cell $i$ in the fluid configuration
$N_\text{f},\mathbf{r}^{N_\text{f}}$. Therefore, the definition of
$n_{ij,\text{dis}}$ for $i=j$ coincides, up to inessential factors,
with the one of the static overlap computed in many simulation studies
of PP fluid systems.
\cite{CavGriVer07PRL,BirBouCavGriVer08NatPhys,CavGriVer10JSMTE,%
  CavGriVer12JCP,GraTroCavGriVer13JCP,KobRolBer12NatPhys,%
  KobRolBer12PhysProc,BerKob12PRE,KobBer13PRL,SauLev11PRL,ChaChaTar12PRL,%
  HocMarRei12PRL,JacFul13PRE,BirKarPro13PRL}

The ensuing result, that the overlap function is simply determined by
the short-range part of the disconnected two-point density, is the
generalization to arbitrary pinning geometries of the one reported in
Ref.~\onlinecite{ChaChaTar13JCP}, where homogeneous PP fluid systems
and a modified overlap function better suited for this case are
considered. \cite{ChaChaTar13JCP,ChaTar13PRE,KarPar13PNAS} It is
hardly spectacular, but has the merit of clearly showing that, for our
purpose, the relevant aspect of the theory of adsorption in disordered
porous media is its analysis of disconnected density correlations.

In this context, one usually deals with the so-called disconnected
total correlation function $h_\text{dis}(\mathbf{x},\mathbf{y})$,
generically defined through
\begin{equation}\label{eq:defhdis}
  \rho_\text{dis}(\mathbf{x},\mathbf{y}) = \rho_\text{f}(\mathbf{x})
  \rho_\text{f}(\mathbf{y}) \left[ h_\text{dis}(\mathbf{x},\mathbf{y})
    + 1 \right].
\end{equation}
This relation becomes particularly simple for the PP systems, since
the density factors are mere constants, in the whole volume of the
homogeneous systems or in the fluid domain of the heterogeneous
ones. Therefore, the overlap function is also straightforwardly
determined by the short-range part of
$h_\text{dis}(\mathbf{x},\mathbf{y})$, that we now investigate.

\subsection{The disconnected total correlation function}

The theory of adsorption in disordered porous solids consists of a
number of developments essentially based on cluster expansions and/or
the use of the replica trick. \cite{MadGla88JSP,Mad92JCP,%
  GivSte92JCP,LomGivSteWeiLev93PRE,GivSte94PA,RosTarSte94JCP,%
  DonKieRos94PRE,Sch02PRE,Sch03PRE,ReiSch04JSP,Sch09PRE,LafCue06PRE}
Regarding the study of the disconnected total correlation function,
most of them have been oriented towards the formulation of integral
equation theories. \footnote{Possibilities also exist in the framework
  of the density functional theory, \cite{Sch02PRE,Sch03PRE,%
    ReiSch04JSP,Sch09PRE,LafCue06PRE} through the use of specially
  tailored test-particle methods, \cite{Sch09PRE} but they have been
  little explored.}

With respect to the PP systems, two models are of particular
relevance.

On the one hand, as pointed out in Ref.~\onlinecite{Kra10PRE}, the
homogeneous PP fluid systems are special cases of the
templated-depleted systems put forward by Van Tassel \textit{et al.}
\cite{TasTalVioTar97PRE,Tas97JCP,Tas99PRE,ZhaTas00JCP,ZhaTas00MP,%
  ZhaCheTas01PRE} Indeed, in this class of models, the matrix samples
are obtained by freezing equilibrium configurations of a binary
mixture and by removing one component acting as a template for the
other. Then, a fluid is introduced in the porous solid made of the
remaining particles. One can easily figure out that, from this point
of view, a homogeneous PP fluid system with pinning fraction $x$ is
nothing but a templated-depleted system built on an ideal binary
mixture with matrix and template number fractions $x$ and $1-x$,
respectively, in which the template particles are reinjected as the
adsorbed fluid.

On the other hand, for the study of fluids adsorbed on substrates
imposing amorphous boundary conditions at their surface, Dong \emph{et
  al.} have considered a generic model in which the procedure for the
preparation of the solid is a heterogeneous pinning process, but no
assumption is made on the fluid in contact with it.
\cite{DonKieRos94PRE} Again, it is clear that the heterogeneous PP
systems are no more than a special instance of this
setup. \footnote{There is no infinite potential excluding the fluid
  from the matrix in the original work reported in
  Ref.~\onlinecite{DonKieRos94PRE}, but one can easily add it to the
  formalism and check that this does not change the equations. For
  definiteness, it should also be mentioned that the species referred
  to as the template in the present work corresponds in
  Ref.~\onlinecite{DonKieRos94PRE} to the ``ghost'' matrix particles
  that do not interact with the fluid.}

For both models, their proposers have derived sets of Ornstein-Zernike
(OZ) equations that must be adapted to the PP fluid systems by using
the relations \eqref{eq:identhomo} and \eqref{eq:identhetero}. In
passing, one may note that in both cases, the last one of these
relations implies that the disconnected total correlation function
does not only describe the sample-to-sample fluctuations of the random
one-body fluid density, as follows from Eqs.~\eqref{eq:discdens} and
\eqref{eq:defhdis}, but also quantifies the memory kept by the fluid
of the positions occupied by the unpinned particles at the time of
preparation of the substrate.

The results are very compact, as they reduce to only two linearly
independent OZ equations. First, the one describing the original bulk
system,
\begin{equation}\label{eq:bulkOZ}
  h(r) = c(r) + \rho c \otimes h(r),
\end{equation}
comes out unchanged, with $\otimes$ denoting a convolution in real
space and $c(r)$ the direct correlation function. This is expected,
since $h(r)$ and $c(r)$ actually have the status of input quantities
for the problem at hand, stemming from a prior study of the structure
of the bulk fluid from which the PP systems are prepared. Second, the
disconnected correlation functions are found to obey
\begin{multline}\label{eq:disOZ}
  h_\text{dis}(r) = c_\text{dis}(r) + \rho c \otimes h(r) \\ - (1-x)
  \rho [c - c_\text{dis}] \otimes [h - h_\text{dis}](r)
\end{multline}
for the homogeneous PP systems, \cite{Kra10PRE} and
\begin{multline}\label{eq:disOZhet}
  h_\text{dis}(\mathbf{x},\mathbf{y}) =
  c_\text{dis}(\mathbf{x},\mathbf{y}) + \rho c \otimes
  h(|\mathbf{x}-\mathbf{y}|) \\ - \rho \int d\mathbf{u} [c -
  c_\text{dis}](\mathbf{x},\mathbf{u}) \chi_{\mathcal{R}}(\mathbf{u})
  [h - h_\text{dis}](\mathbf{u},\mathbf{y})
\end{multline}
for the heterogeneous ones, with $c_\text{dis}(r)$
[$c_\text{dis}(\mathbf{x},\mathbf{y})$] the disconnected direct
correlation function associated with $h_\text{dis}(r)$
[$h_\text{dis}(\mathbf{x},\mathbf{y})$].  Using
$\chi_{\mathcal{R}}(\mathbf{u}) = 1 -
\chi_{\overline{\mathcal{R}}}(\mathbf{u})$, a structural similarity
between both equations immediately appears, with
$\chi_{\overline{\mathcal{R}}}(\mathbf{u})$ playing the role of a
space-dependent pinning fraction. With Eq.~\eqref{eq:disOZhet}, one
can also remark the unusual situation of an inhomogeneous integral
equation theory that does not require a computation of the density
profile through an additional relation. \cite{[][{; see Chap. 9 for
    the asymptotic behavior of the correlation functions in simple
    fluids, and Chap. 11 for the inhomogeneous integral equation
    theories.}]Att02book}

Further results of interest can be obtained by delving into the
peculiar diagrammatic structure of the underlying cluster expansions.
\cite{MadGla88JSP,Mad92JCP,GivSte92JCP,LomGivSteWeiLev93PRE,GivSte94PA,%
  DonKieRos94PRE}

From a general point of view, Madden's formalism dealing with
arbitrary fluid-matrix ensembles shows that the diagrams contributing
to the disconnected total correlation function involve many-body
matrix distribution functions of all
orders. \cite{Mad92JCP,Mad95JCPcom} Since in the PP systems these
distribution functions coincide with those of the original bulk fluid,
this observation nicely embodies the characterization of
$h_\text{dis}$ as a point-to-set correlation function probing the full
many-body structure of this fluid. However, this aspect does not
appear to play any particular role as the theory unfolds, so that the
PP systems do not seem to occupy any special place in this framework.

The low pinning fraction regime of the homogeneous PP systems has
recently been considered, with calculations of the leading order
contribution to $h_\text{dis}(r)$ for small $x$.
\cite{ChaChaTar13JCP,SzaFle13EPL} This result can be derived in the
present framework with the help of the generic topological constraints
on the diagrams contributing to $h_\text{dis}(r)$ and
$c_\text{dis}(r)$ for any matrix ensemble. Indeed, they impose that,
in the limit of the matrix density $\rho_\text{m}$ going to zero,
these functions are of order $\rho_\text{m}$ and $\rho_\text{m}^2$,
respectively. \footnote{The key aspect is that, in the diagrams
  contributing to the disconnected correlation functions, all paths
  linking the fluid root points must go through at least one matrix
  field point. Since nodal points are allowed in $h_\text{dis}$, it is
  possible to have only one matrix field point, which is precisely and
  necessarily nodal. On the other hand, nodal points being forbidden
  in $c_\text{dis}$, at least two matrix field points must be
  present.} Plugging this into Eq.~\eqref{eq:disOZ}, knowing that
$\rho_\text{m}$ means $x \rho$ and that $h(r)$ and $c(r)$ are of order
$1$ (remember that they simply refer to the bulk fluid from which the
PP system is prepared), leads after simple manipulations to
\begin{equation} \label{eq:asympt} h_\text{dis}(r) = x \rho h \otimes
  h(r), \qquad x \rho \to 0,
\end{equation}
a result that can be confirmed independently of Eq.~\eqref{eq:disOZ}
by a direct inspection of the diagrams contributing to
$h_\text{dis}(r)$ at order $\rho_\text{m}$. If the limit $x\rho\to0$
involves $\rho\to0$, i.e., if homogeneous pinning in an extremely
dilute gas is considered, $h(r)$ in this equation implicitly stands
for $\lim_{\rho\to0} h(r)$.

In the limit $x\to0$ with $\rho$ fixed, Eq.~\eqref{eq:asympt}
expresses the linear response of the fluid to the homogeneous pinning
constraint. Its derivation in Refs.~\onlinecite{ChaChaTar13JCP} and
\onlinecite{SzaFle13EPL} has been through \emph{ad hoc} calculations
especially tailored for the study of PP systems. By considering the
problem in the general settings of the theory of adsorption, the
present approach has the advantage that it lends itself to simple
generalizations. For instance, the above-mentioned diagrammatic
inspection of $h_\text{dis}(r)$ shows that Eq.~\eqref{eq:asympt}, with
$x\rho$ replaced by $\rho_\text{m}$, generically describes the
response of a fluid to a vanishingly small amount of randomly placed
matrix particles identical to the fluid ones, irrespective of the
detailed disordered structure of the matrix. Thus, not only has the
contribution to $h_\text{dis}(r)$ reported in Eq.~\eqref{eq:asympt}
little to do with point-to-set correlations, as argued in
Refs.~\onlinecite{ChaChaTar13JCP} and \onlinecite{SzaFle13EPL}, but it
is also completely unspecific of the PP systems.

The question of an analogous contribution in the heterogeneous case
has been raised by some authors.
\cite{SzaFle13EPL,GraTroCavGriVer13JCP} It can be addressed along
exactly the same lines as above, thanks to the cluster expansions of
Ref.~\onlinecite{DonKieRos94PRE}. Here, the asymptotic regime in which
the perturbation due to the pinning is small corresponds to locations
in $\mathcal{R}$ far from any fluid-solid interface. The result is
derived in the Appendix and reads
\begin{multline}\label{eq:asymhet}
  h_\text{dis}(\mathbf{x},\mathbf{y}) = \rho \int d\mathbf{u}
  h(|\mathbf{x}-\mathbf{u}|) \chi_{\overline{\mathcal{R}}}(\mathbf{u})
  h(|\mathbf{u}-\mathbf{y}|), \\ \mathbf{x},\mathbf{y} \in
  \mathcal{R}, d(\mathbf{x},\overline{\mathcal{R}}) \text{ and/or }
  d(\mathbf{y},\overline{\mathcal{R}}) \to+\infty,
\end{multline}
with $d(\mathbf{x},\overline{\mathcal{R}}) =
\min_{\mathbf{r}\in\overline{\mathcal{R}}} |\mathbf{x}-\mathbf{r}|$.
The above-mentioned structural similarity between
Eqs.~\eqref{eq:disOZ} and \eqref{eq:disOZhet} remains between
Eqs.~\eqref{eq:asympt} and \eqref{eq:asymhet}.

Assuming that the separate problem of the structure of the original
bulk fluid is under control, an explicit calculation of $h_\text{dis}$
beyond these asymptotic regimes in the framework of an integral
equation theory requires an approximate closure of
Eqs.~\eqref{eq:disOZ} or \eqref{eq:disOZhet}, in the form of an
additional independent relation involving $c_\text{dis}$ and
$h_\text{dis}$. A very convenient route towards such closures is the
use of the replica trick. \cite{GivSte92JCP,LomGivSteWeiLev93PRE,%
  GivSte94PA,RosTarSte94JCP}

In order to illustrate such a scheme with a simple numerical
computation, we consider the homogeneous PP fluid systems prepared
from the one-component hard-sphere fluid, in the so-called replica
Percus-Yevick (PY) approximation. \cite{GivSte92JCP,%
  LomGivSteWeiLev93PRE,GivSte94PA,ZhaTas00JCP,ZhaTas00MP} This closure
implies that the structure of the bulk fluid is handled through the
usual PY approximation, with a well-known analytic solution for the
hard-sphere fluid, \cite{macdohansen3ed} and that $c_\text{dis}(r)$ is
identically set to zero.

\begin{figure*}
\includegraphics*{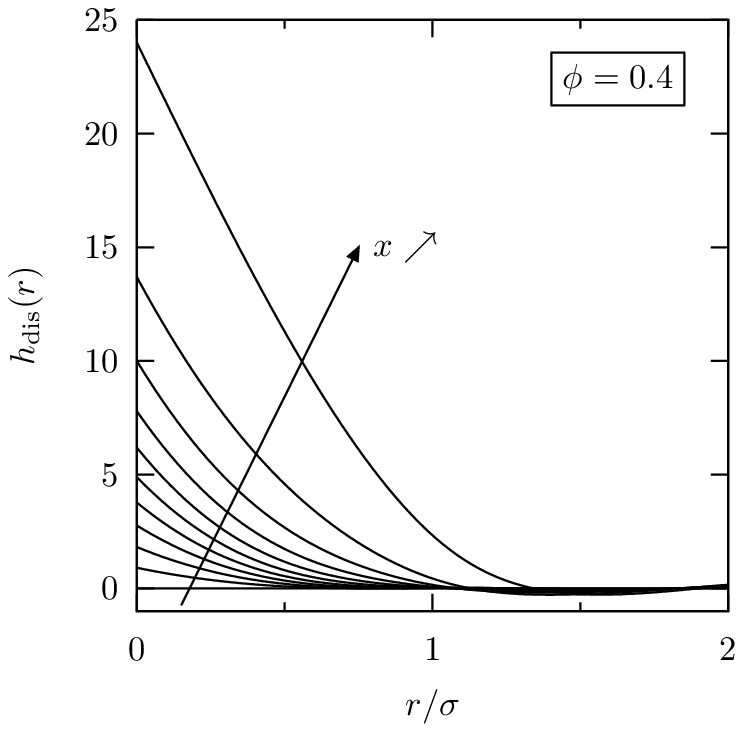}\hspace{2em}\includegraphics*{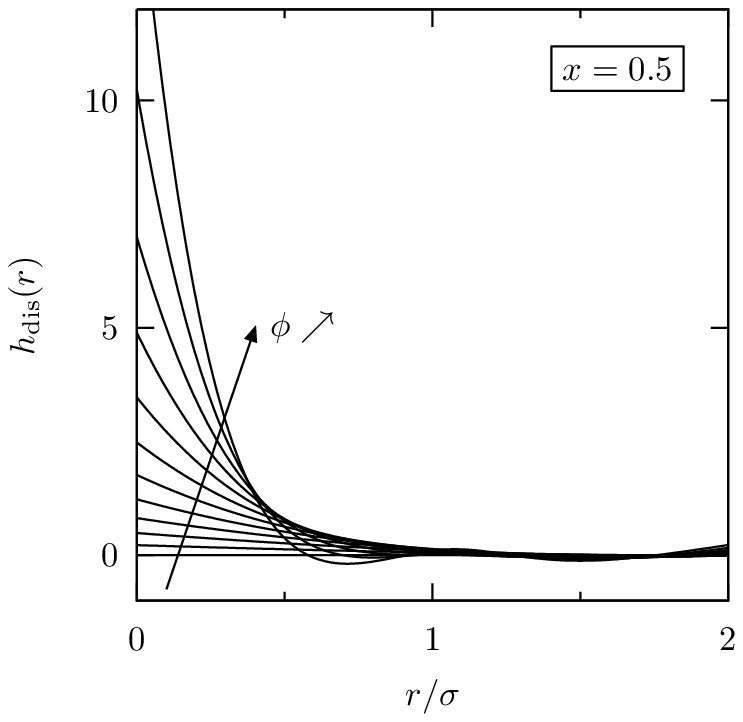}
\caption{\label{fighdvs} Disconnected total correlation function
  $h_\text{dis}(r)$ of the homogeneous partly pinned hard-sphere fluid
  system, computed in the replica Percus-Yevick
  approximation. $\sigma$ denotes the particle diameter, $\phi$ the
  compacity of the original bulk system, and $x$ the pinning
  fraction. Left: $h_\text{dis}(r)$ as a function of $x$
  ($x=0,0.1,0.2,\ldots,1$) for $\phi=0.4$. Right: $h_\text{dis}(r)$ as
  a function of $\phi$ ($\phi=0,0.05,0.1,\ldots,0.55$) for $x=0.5$.}
\end{figure*}

The resulting disconnected total correlation function
$h_\text{dis}(r)$ is displayed in Fig.~\ref{fighdvs} for
representative values of the compacity of the original bulk system
$\phi=\rho\pi\sigma^3/6$ ($\sigma$ denotes the particle diameter) and
of the pinning fraction $x$. There, $h_\text{dis}(r)$ is seen to have
a perfectly regular short-distance behavior and to change smoothly
with the parameters defining the PP system. Therefore, one can use its
value at $r=0$ as a direct proxy for the variations of the static
overlap function.

\begin{figure}
\includegraphics*{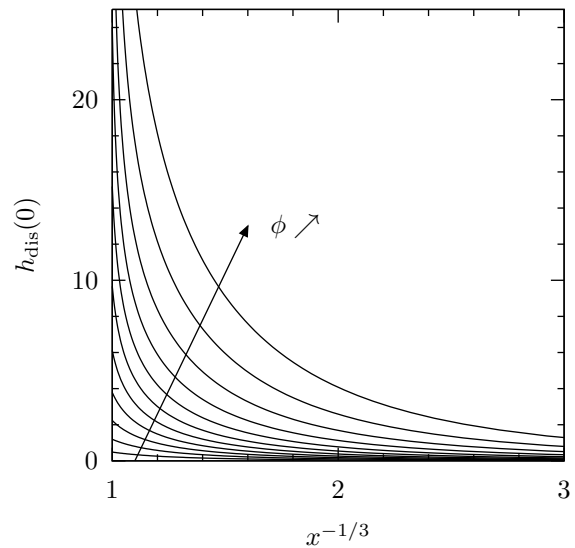}
\caption{\label{figzero} Dependence on the pinning fraction $x$ of the
  zero-separation value $h_\text{dis}(0)$ of the disconnected total
  correlation function of the homogeneous partly pinned hard-sphere
  fluid system, computed in the replica Percus-Yevick
  approximation. The compacities of the original bulk systems are
  $\phi=0.05,0.1,\ldots,0.55$.}
\end{figure}

Accordingly, the $x$-dependence of $h_\text{dis}(0)$ is shown in
Fig.~\ref{figzero} for different values of $\phi$. Here, we follow
Ref.~\onlinecite{ChaChaTar12PRL} and use the variable $x^{-1/3}$, as a
reflection of the $(x \rho)^{-1/3}$ scaling of the typical distance
between pinned particles. \footnote{In the more recent
  Refs.~\onlinecite{ChaChaTar13JCP} and \onlinecite{ChaTar13PRE}, the
  full length $(x \rho)^{-1/3}$ is used as the variable in similar
  plots. We do not do so here, because the resulting horizontal shift
  with $\rho$ of the endpoints corresponding to $x=1$ leads to
  overlapping curves and a less legible figure.} The replica PY
$h_\text{dis}(0)$ is clearly seen to reproduce the qualitative
behavior of the overlap function as measured in computer simulations
of dense hard-sphere mixtures, in particular, its rapid increase when
$x$ goes to one. \cite{ChaChaTar12PRL,ChaChaTar13JCP,ChaTar13PRE}
Therefore, as in the simulation studies, it is possible with the
present theory to pinpoint a moderately increasing density-dependent
static lengthscale, by recording as a function of $\rho$ the value of
$(x \rho)^{-1/3}$ at which $h_\text{dis}(0)$ reaches some intermediate
threshold value, e.g., $h_\text{dis}(0)=10$, when $x$ increases. This
lengthscale cannot have any connection with glassiness, that is not
captured by this theory (on the inability of the PY approximation to
give a glass transition, see the appendix of
Ref.~\onlinecite{CarFraPar99JCP}). In passing, regarding the large $x$
regime, it might be mentioned that an analytic expression for the
terminal value of $h_\text{dis}(0)$ at $x=1$ is provided by the
present closure,
\begin{equation}
  h_\text{dis}(0) = \frac{(1+2\phi)^2}{(1-\phi)^4} - 1, \qquad x=
  1.
\end{equation}

\begin{figure}
\includegraphics*{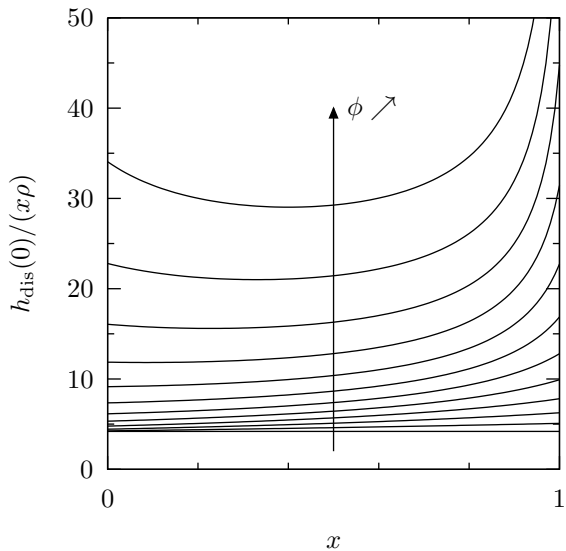}
\caption{\label{figzeronl} Dependence on the pinning fraction $x$ of
  the zero-separation value $h_\text{dis}(0)$ of the disconnected
  total correlation function of the homogeneous partly pinned
  hard-sphere fluid system, computed in the replica Percus-Yevick
  approximation. In order to magnify the departure from the linear
  response result, Eq.~\eqref{eq:asympt}, with increasing $x$,
  $h_\text{dis}(0)$ is divided by $x \rho$. The compacities of the
  original bulk systems are $\phi=0,0.05,0.1,\ldots,0.55$.}
\end{figure}

We close this part with Fig.~\ref{figzeronl}, where the $x$-dependence
of the replica PY $h_\text{dis}(0)$ is reported in a way that
magnifies the departure from the linear response behavior,
Eq.~\eqref{eq:asympt}, when $x$ grows from zero. At fixed
$\phi\lesssim0.4$, $h_\text{dis}(0)/(x\rho)$ is found to be an
increasing function of $x$. This means in particular that the
nonlinear contribution to $h_\text{dis}(0)$ is always positive in this
low density regime. On the other hand, at fixed $\phi\gtrsim0.4$,
$h_\text{dis}(0)/(x\rho)$ first decreases then increases with $x$,
hence develops a minimum that is more and more marked when $\phi$
grows. Therefore, in this moderate-to-high density regime, there is a
sign change of the nonlinear contribution to $h_\text{dis}(0)$, from
negative at low $x$ to positive at high $x$. Obviously, these results
should be taken with care because of the approximations underlying the
replica PY closure, the neglect of $c_\text{dis}(r)$ of order $(x
\rho)^2$, in particular. However, they should also serve as a reminder
that, in principle, the behavior of the disconnected total correlation
function, hence of the overlap, can undergo nontrivial qualitative
changes unrelated to the physics of the glass transition.

\subsection{Discussion}

We now contemplate the possible relevance of the above developments
for the study of the PP systems prepared from glassforming liquids.

As long as the systems under consideration behave ergodically, i.e.,
as long as the global statistical-mechanical framework of
Sec.~\ref{sec:basic} is appropriate, everything that has been said
before remains valid. One deals with the same intricate measure of the
interplay between the evolutions of the (many-body) correlations in a
bulk fluid and its response to an externally imposed quenched
disorder. Because of the lower temperatures and/or higher densities,
the theoretical issue surely becomes more challenging, with
uncertainties on the validity of the usual simple approximations,
\cite{macdohansen3ed} but the basic working equations such as the
above OZ relations apply unmodified.

The situation changes when the possibility of broken ergodicity is
introduced. This is the case in studies of PP systems whose aim is to
probe a putative complex coarse-grained free-energy landscape
characterizing the original bulk glassforming liquid and the related
concept of an incipient amorphous order. \cite{BouBir04JCP,%
  CavGriVer07PRL,BirBouCavGriVer08NatPhys,GraTroCavGriVer13JCP}
Indeed, one of their central tenets is that a generic outgrowth of
such a landscape will be the existence of many (meta)stable states
accessible to the confined fluid.

One can easily devise how the previous formalism is altered in this
case, borrowing ideas from the mean-field theory of
glasses. \cite{ParZam10RMP,FraSem11book} To shorten the equations, we
write $(N_\text{m}, \mathbf{q}^{N_\text{m}})\equiv\mathbf{m}$ and
$(N_\text{f},\mathbf{r}^{N_\text{f}})\equiv\mathbf{f}$. The relevant
random density field becomes state-dependent in addition to its
realization-dependence and reads
\begin{equation}
  \rho^\alpha_\text{f}[\mathbf{m}](\mathbf{x})
  = \left\langle \hat{\rho}_\text{f}(\mathbf{x};\mathbf{f})
  \right\rangle_{\mathbf{m}}^\alpha,
\end{equation}
where $\langle \cdots \rangle_{\mathbf{m}}^\alpha$ denotes the
realization-dependent thermal average inside state $\alpha$. Its
covariance defines a glassy disconnected two-point density,
\begin{equation}
  \rho^\text{g}_\text{dis}(\mathbf{x},\mathbf{y}) =
  \overline{\{\rho^\alpha_\text{f}[\mathbf{m}](\mathbf{x})
    \rho^\alpha_\text{f}[\mathbf{m}](\mathbf{y})\}_{\mathbf{m}}},
\end{equation}
where $\{\cdots\}_{\mathbf{m}}$ represents the realization-dependent
thermodynamically weighted average over the states and
$\overline{\cdots}$ still is the average over the matrix
realizations. This expression can be reorganized by defining
\begin{equation}
  \delta \rho^\alpha_\text{f}[\mathbf{m}](\mathbf{x}) =
  \rho^\alpha_\text{f}[\mathbf{m}](\mathbf{x})
  - \{\rho^\alpha_\text{f}[\mathbf{m}](\mathbf{x}) \}_{\mathbf{m}},
\end{equation}
so that
\begin{multline} \label{eq:multistate}
  \rho^\text{g}_\text{dis}(\mathbf{x},\mathbf{y}) = \overline{\{
    \rho^\alpha_\text{f}[\mathbf{m}](\mathbf{x}) \}_{\mathbf{m}} \{
    \rho^\alpha_\text{f}[\mathbf{m}](\mathbf{y})\}_{\mathbf{m}}} \\ +
  \overline{ \{ \delta\rho^\alpha_\text{f}[\mathbf{m}](\mathbf{x})
    \delta\rho^\alpha_\text{f}[\mathbf{m}](\mathbf{y})
    \}_{\mathbf{m}}}.
\end{multline}
When the equilibrium consists of a collection of many metastable
states with a finite configurational entropy, a remarkable feature of
the mean-field theory leads to the equality $\{
\rho^\alpha_\text{f}[\mathbf{m}](\mathbf{x}) \}_{\mathbf{m}} =
\rho_\text{f}[\mathbf{m}](\mathbf{x})$, i.e., the ergodic density
profile in a given sample is recovered through state averaging. The
first term in Eq.~\eqref{eq:multistate} is therefore
$\rho_\text{dis}(\mathbf{x},\mathbf{y})$, and one gets
\begin{equation} \label{eq:multistate2}
  \rho^\text{g}_\text{dis}(\mathbf{x},\mathbf{y}) =
  \rho_\text{dis}(\mathbf{x},\mathbf{y}) +
  \delta\rho^\text{g}_\text{dis}(\mathbf{x},\mathbf{y}),
\end{equation}
where $\delta\rho^\text{g}_\text{dis}(\mathbf{x},\mathbf{y}) =
\overline{ \{ \delta\rho^\alpha_\text{f}[\mathbf{m}](\mathbf{x})
  \delta\rho^\alpha_\text{f}[\mathbf{m}](\mathbf{y}) \}_{\mathbf{m}}}$
is a kind of Edwards-Anderson or nonergodicity parameter for the PP
system. By setting $\mathbf{x}=\mathbf{y}=\mathbf{x}_0$, where
$\mathbf{x}_0$ is some convenient reference point, one obtains a
\emph{bona fide} point-to-set correlation function,
$\rho^\text{g}_\text{dis}(\mathbf{x}_0,\mathbf{x}_0)$, essentially
equivalent to the overlap, which would simply correspond to an
integral of $\rho^\text{g}_\text{dis}(\mathbf{x},\mathbf{y})$ over a
small volume around $\mathbf{x}_0$.

From Eq.~\eqref{eq:multistate2}, one immediately sees that, if there
exists a contribution to the point-to-set correlation function
originating in genuine glassy effects, it will typically come
aggregated with a more commonplace one merely reflecting the
(hypothetical) response of a normal fluid to the pinning disorder. On
general grounds (see also Fig.~\ref{figzero}), the latter contribution
can be expected to grow stronger and longer-ranged as the density of
the fluid increases and/or its temperature decreases, and, as stressed
in the previous section with Fig.~\ref{figzeronl}, although the
underlying physics is somewhat simple, its behavior is not necessarily
so. Such features might obscure or hide evolutions of the glassy
contribution, which is usually the actual quantity of interest [for
instance, the fact that the derivation of rigorous bounds between
length and time scales requires the existence of some kind of
Edwards-Anderson parameter \cite{MonSem06JSP_2,MonSem06JSP,%
  FraSem11book} clearly points towards a preeminent role of
$\delta\rho^\text{g}_\text{dis}(\mathbf{x}_0,\mathbf{x}_0)$].

These observations suggest ways in which the toolbox of the theory of
adsorption on random substrates could be put to good use for the study
of glassy PP systems. For instance, one could try to combine
simulation data and theoretical predictions to correct the measured
point-to-set correlation functions for the direct effect of
adsorption, i.e., in terms of disconnected two-point densities,
replace the analysis of
$\rho^\text{g}_\text{dis}(\mathbf{x}_0,\mathbf{x}_0)$ by that of
$\rho^\text{g}_\text{dis}(\mathbf{x}_0,\mathbf{x}_0) -
\rho_\text{dis}(\mathbf{x}_0,\mathbf{x}_0)$. This could well turn out
an unnecessary complication, if the glassy contribution is eventually
shown to dominate, but it is also possible that no evidence for
something like
$\delta\rho^\text{g}_\text{dis}(\mathbf{x}_0,\mathbf{x}_0)$ is finally
found.  In any case, the current approach, in which no attempt is made
to resolve two physically distinct contributions to the overlap, might
not be optimal for sharp characterizations of the putative amorphous
order developing in glassy systems.
\section{The homogeneous partly pinned lattice gas}
\label{sec:lattice}

In this Section, we consider the outcome of the homogeneous pinning
procedure when applied to a simple lattice gas, its equivalence with a
peculiar instance of the random field Ising model (RFIM), and some
possible consequences of this mapping.

\subsection{Homogeneous pinning process in a lattice gas and mapping
  onto a random field Ising model}

The pinning process described in Sec.~\ref{subsec:reminder} applies
equally well if the particle positions are restricted to the vertices
of a periodic lattice, i.e., if the simple fluid under consideration
is actually a lattice-gas model. Thus, we might investigate along the
same lines the most common member of this family, characterized by an
infinite on-site exclusion preventing multiple occupancy of a lattice
vertex and an attractive pair interaction $-w$ ($w>0$) between
particles located on nearest-neighbor sites.

As is well known, there are two ways to encode the configurations of
such a lattice model. The first one is by simply keeping track of the
particle positions, as with any other fluid model. The second one is
specifically lattice-based and records, for each site $i$ of the
lattice, the number $n_i$ of particles sitting on this site. For
systems with infinite on-site exclusion, this occupancy number is a
binary variable, $n_i=0$ if the site is empty, $n_i=1$ if it is
occupied, easily mapped onto an Ising spin. Our first task is
therefore to rephrase the results expressed in terms of particle
positions in the language of occupancy variables.

The results for the bulk are very classic. \cite{LavBel99bookv1} In
terms of the variables $(n_1,n_2,\ldots,n_V) \equiv \mathbf{n}^V$ ($V$
now denotes the total number of sites), the grand-canonical
statistical mechanics of the lattice gas is ruled by the effective
Hamiltonian
\begin{equation}
  H_\text{bulk}(\mathbf{n}^V) = - w \sum_{\langle i j \rangle}
  n_i n_j - \mu \sum_i n_i,
\end{equation}
where $\langle i j \rangle$ generically denotes the nearest-neighbor
pairs of sites, to which the first sum is restricted, and the chemical
potential $\mu = k_\text{B}T \ln z$ has been introduced. Accordingly,
the probability of a configuration $\mathbf{n}^V$ reads
\begin{equation}
  \mathcal{P}_\text{bulk}(\mathbf{n}^V) = \frac{e^{-\beta
      H_\text{bulk}(\mathbf{n}^V)}}{\Xi_\text{bulk}},
  \
  \Xi_\text{bulk} = \sum_{\mathbf{n}^V} e^{-\beta
    H_\text{bulk}(\mathbf{n}^V)}.
\end{equation}

The pinning process amounts to randomly choosing a fraction of the
sites that are occupied in an instantaneous bulk configuration, to
impose that they will remain so forever. This divides the lattice
vertices into two complementary groups, those permanently hosting a
matrix particle and those remaining accessible to the fluid particles,
identified by the sets of indices $\{i_\text{m}\}$ and
$\{i_\text{f}\}$, respectively. Therefore, a configuration
$\mathbf{n}^V$ after the pinning step naturally decomposes as
$\mathbf{n}^{\{i_\text{m}\}} \oplus \mathbf{n}^{\{i_\text{f}\}}$,
where $\mathbf{n}^{\{i_\text{m}\}} \equiv (n_i)_{i\in\{i_\text{m}\}}$
and $\mathbf{n}^{\{i_\text{f}\}} \equiv (n_i)_{i\in\{i_\text{f}\}}$,
knowing that $n_i=1$ for all $i\in\{i_\text{m}\}$ by
construction. Then, noting that $N_\text{m} =
\sum_{i\in\{i_\text{m}\}} n_i$ and $N_\text{t} =
\sum_{i\in\{i_\text{f}\}} n_i$, the joint probability that a matrix
realization labeled by $\{i_\text{m}\}$ is generated while the
occupancy vector of the remaining sites is
$\mathbf{n}^{\{i_\text{f}\}}$ follows from Eq.~\eqref{pinjoint} as
\begin{multline} \label{lgpinjoint}
  \mathcal{P}_\text{mt}(\{i_\text{m}\},
  \mathbf{n}^{\{i_\text{f}\}}) = \\
  \frac{e^{ -\beta H_\text{bulk}(\mathbf{n}^{\{i_\text{m}\}} \oplus
      \mathbf{n}^{\{i_\text{f}\}}) + \ln(x) \sum_{i\in\{i_\text{m}\}}
      n_i + \ln(1-x) \sum_{i\in\{i_\text{f}\}} n_i}}{\Xi_\text{bulk}},
\end{multline}
from which the overall probability of the matrix configuration is
obtained,
\begin{equation} \label{lgpinmatrix}
  \mathcal{P}_\text{m}(\{i_\text{m}\}) =
  \sum_{\mathbf{n}^{\{i_\text{f}\}}}
  \mathcal{P}_\text{mt}(\{i_\text{m}\},\mathbf{n}^{\{i_\text{f}\}}).
\end{equation}

The effective Hamiltonian describing the confined fluid in a given
matrix realization $\{i_\text{m}\}$ in terms of the variables
$\mathbf{n}^{\{i_\text{f}\}}$ is retrieved from the argument of the
exponential in Eq.~\eqref{lgpinjoint} by removing all the pure matrix
contributions, so that one gets
\begin{multline} 
  H_\text{f}(\mathbf{n}^{\{i_\text{f}\}}|\{i_\text{m}\}) = - w
  \sum_{\substack{\langle i j \rangle \\ i,j \in\{i_\text{f}\}}} n_i
  n_j - w \!\!\!\!\!\!\sum_{\substack{\langle i j \rangle \\
      i\in\{i_\text{f}\}, j\in\{i_\text{m}\}}}\!\!\!\!\!\! n_i n_j \\
  - w \!\!\!\!\!\!\sum_{\substack{\langle i j \rangle \\
      i\in\{i_\text{m}\}, j\in\{i_\text{f}\}}}\!\!\!\!\!\! n_i n_j -
  \left[ \mu + k_\text{B}T \ln(1-x) \right] \sum_{i\in\{i_\text{f}\}}
  n_i,
\end{multline}
where the condition $n_i=1$ for $i\in\{i_\text{m}\}$ has not been
explicitly used for later convenience.

In order to lift the restrictions on the sums in $H_\text{f}$, we take
advantage of the fact that the physical properties of the system are
left unaffected if a term independent of the fluid variables
$\mathbf{n}^{\{i_\text{f}\}}$ is added to the
Hamiltonian. \footnote{Another possibility is to introduce matrix
  occupancy variables, $(u_1,u_2,\ldots,u_V)$, such that $u_i=1$ if
  $i\in\{i_\text{m}\}$ and $u_i=0$ otherwise. One then obtains the
  Hamiltonian 
  \begin{multline}
    H_\text{f}(\mathbf{n}^{V}|\mathbf{u}^{V}) = \\ - w \sum_{\langle i
      j \rangle} n_i n_j (1-u_i) (1-u_j) + n_i (1-u_i) u_j + n_j
    (1-u_j) u_i \\ - \left[ \mu + k_\text{B}T \ln(1-x) \right]
    \sum_{i} n_i (1-u_i).
  \end{multline}
  This formulation shows that the present PP system bears strong
  connection to a lattice-gas model previously studied by Rosinberg
  \emph{et al.}, who showed that it can be readily transformed into an
  Ising model with correlated site dilution and random
  fields. \cite{PitRosSteTar95PRL,PitRosTar96MS,KieRosTar97JSP,
    KieRosTarPit98MP}} Thus, one can introduce the pure matrix
contribution $ - w \sum_{\langle i j \rangle; i,j \in\{i_\text{m}\}}
n_i n_j - \left[ \mu + k_\text{B}T \ln(1-x) \right]
\sum_{i\in\{i_\text{m}\}} n_i$, consisting of the matrix-matrix
interactions at the time of pinning plus a suitable chemical potential
term, and redefine
\begin{multline} 
  H_\text{f}(\mathbf{n}^{\{i_\text{f}\}}|\{i_\text{m}\}) = \\
  - w \sum_{\langle i j \rangle} n_i n_j - \left[ \mu + k_\text{B}T
    \ln(1-x) \right] \sum_i n_i.
\end{multline}
Note that this differs from the argument of the exponential in
Eq.~\eqref{lgpinjoint}. This expression is in line with the picture of
the PP systems as constrained systems, since the above Hamiltonian is
the one of a mere lattice gas, in which occupancy of some sites is
imposed.

Ising spin variables $\sigma_i=\pm1$ are then readily introduced
through the simple transformation $n_i = (1+\sigma_i)/2$, which maps
$H_\text{f}(\mathbf{n}^{\{i_\text{f}\}}|\{i_\text{m}\})$ onto
\begin{multline} \label{eq:pinnedising}
  H_\text{s}(\boldsymbol{\sigma}^{\{i_\text{f}\}}|\{i_\text{m}\}) = \\
  - \frac{w}{4} \sum_{\langle i j \rangle} \sigma_i \sigma_j  -
  \frac12 \left[ \mu + \frac{q w}{2} + k_\text{B}T \ln(1-x) \right]
  \sum_i \sigma_i.
\end{multline}
where $q$ denotes the coordination number and irrelevant
spin-independent contributions are omitted.

This Hamiltonian describes Ising variables evolving in the presence of
permanently frozen up spins, $\sigma_i=+1$ for
$i\in\{i_\text{m}\}$. Up to now, this freezing has been imposed by
hand. However, it can be made a consequence of the Hamiltonian of the
system, if an infinite on-site magnetic field forcing the spins to
align is introduced, an idea that goes back to Grinstein and Mukamel
in the context of the RFIM. \cite{GriMuk83PRB} So, the present PP
lattice gas is eventually shown to be equivalent to a RFIM, with the
usual Hamiltonian
\begin{equation} \label{eq:RFIM}
  H_\text{RFIM}(\boldsymbol{\sigma}^{V}|\mathbf{h}^{V}) =
  - \frac{w}{4} \sum_{\langle i j \rangle} \sigma_i \sigma_j -
  \sum_i h_i \sigma_i,
\end{equation}
and a random field realization
\begin{equation} \label{eq:randomfield}
  h_i = 
  \begin{cases}
    \dfrac12 \left[ \mu + \dfrac{q w}{2} + k_\text{B}T \ln(1-x) \right]
    & \mbox{if $i\in\{i_\text{f}\}$,} \\
    +\infty & \mbox{if $i\in\{i_\text{m}\}$.}
    \end{cases}
\end{equation}

In order to consider the mapping as complete, one might also wish to
express the probability distribution of the random field
configurations, as characterized by the sets of indices
$\{i_\text{m}\}$, in terms of Ising spins. This only requires a
rewriting of Eq.~\eqref{lgpinmatrix}, leading to
\begin{multline}
  \mathcal{P}_\text{m}(\{i_\text{m}\}) = 
  \sum_{\boldsymbol{\sigma}^{V}} \Big[ \Big(
  \prod\nolimits_{i\in\{i_\text{m}\}} \delta_{+1,\sigma_i} \Big) \\
  \times x^{\sum_{i\in\{i_\text{m}\}} \frac{1+\sigma_i}{2}}
  (1-x)^{\sum_{i\in\{i_\text{f}\}} \frac{1+\sigma_i}{2}}
  \mathcal{P}_\text{Ising}(\boldsymbol{\sigma}^V) \Big],
\end{multline}
with
\begin{equation}
  \mathcal{P}_\text{Ising}(\boldsymbol{\sigma}^V) = \frac{e^{-\beta
      H_\text{Ising}(\boldsymbol{\sigma}^V)}}{\Xi_\text{Ising}},
  \
  \Xi_\text{Ising} = \sum_{\boldsymbol{\sigma}^V} e^{-\beta
    H_\text{Ising}(\boldsymbol{\sigma}^V)},
\end{equation}
where $H_\text{Ising}(\boldsymbol{\sigma}^V)$ is the Ising Hamiltonian
corresponding to $H_\text{bulk}(\mathbf{n}^V)$,\cite{LeeYan52PR}
\begin{equation} 
  H_\text{Ising}(\boldsymbol{\sigma}^{V}) =
  - \frac{w}{4} \sum_{\langle i j \rangle} \sigma_i \sigma_j - \frac12
  \left[ \mu + \frac{q w}{2} \right] \sum_i \sigma_i.
\end{equation}

Obviously, because of the direct correspondence between occupied cells
($n_i=1$) and up spins ($\sigma_i=+1$), the same instance of the RFIM
would be readily obtained if one would start with $H_\text{Ising}$ and
consider a pinning process acting on up spins only.

\subsection{Discussion}

A few comments can be made, based on these results.

A random pinning process in a particle system is found to be
essentially equivalent to a spin pinning process restricted to up
spins in an Ising model. As shown in Ref.~\onlinecite{Kra10PRE}, the
former is accompanied by a change in activity and chemical potential,
from $z$ and $\mu$ for the original bulk fluid to $z(1-x)$ and
$\mu+k_\text{B}T\ln(1-x)$ for the confined fluid, that originates in
simple combinatorics. \footnote{Incidentally, this result implies that
  a homogeneous PP system is never in equilibrium with the bulk fluid
  from which it has been prepared. If the two are brought into
  contact, particles will spontaneously move from the bulk to the
  porous matrix, until equalization of the chemical potentials is
  achieved.}  This translates in the latter into a homogeneous shift
of the external magnetic field in which the (unpinned) spins are
plunged. This shift, equal to $k_\text{B}T \ln(1-x)/2$ and therefore
negative, can be understood as a depolarizing field compensating for
the fraction of pinned up spins. It is required to maintain the same
average configurational properties in the bulk and pinned spin
systems.

This fluid-like pinning scheme in the Ising model is clearly different
from the usual obvious one, in which spins are frozen at random
locations independently of their instantaneous value and no change of
the external magnetic field is required. One can also map the latter
onto a particle pinning scheme, through the correspondence between the
Ising model and a lattice binary mixture. \cite{LavBel99bookv1}
However, in this case, the fluid model is incompressible and the
pinning constraint acts on concentration rather than density
fluctuations, so that one expects distinct physical consequences from
this pinning scenario.

These observations raise the issue of the choice of a pinning scheme
when one wants to investigate the effect of pinned particles on a
fluid on the basis of a phenomenological model that can be very
different from a fluid. Indeed, as found here for the simple Ising
model, there might be more than just one possible pinning scheme, and
choosing any particular one might not be innocent, since not all these
schemes might be representative of an actual pinning process in a
fluid system. In the absence of a guiding principle such as the above
Ising model/lattice gas duality, a conservative strategy could then be
to compare several schemes. To give a concrete example, Jack and
Berthier \cite{JacBer12PRE} have investigated the effect of simple
spin pinning on plaquette models that also admit a dual representation
in terms of defect variables. It would maybe be interesting to
consider the consequences of defect pinning as well.

Another way to write Eqs.~\eqref{eq:RFIM} and \eqref{eq:randomfield}
is
\begin{gather}
  H_\text{RFIM}(\boldsymbol{\sigma}^{V}|\mathbf{h'}^{V}) =
  H_\text{Ising}(\boldsymbol{\sigma}^{V}) - \sum_i h'_i \sigma_i, \\
  h'_i =
  \begin{cases}
    \dfrac12 k_\text{B}T \ln(1-x) & \mbox{if $i\in\{i_\text{f}\}$,} \\
    +\infty & \mbox{if $i\in\{i_\text{m}\}$,}
  \end{cases}
\end{gather}
where the Hamiltonian of the original bulk spin system appears
perturbed by a random field term stemming from the pinning process. It
turns out that this random contribution is very similar to the one
seen in a RFIM built by Cammarota and Biroli in the framework of the
RFOT theory. \cite{CamBir12PNAS,CamBir13JCP} Both are indeed
characterized by a homogeneous negative magnetic field, except at
randomly placed sites where saturating positive fields fully polarize
the spins in the up state.

In most respects, a direct comparison of these two results does not
make sense, because of their very different theoretical
foundations. The computer simulations performed by Cammarota and
Biroli in order to benchmark the behavior of their model
\cite{CamBir13JCP} might however represent an exception, because they
are largely independent from the path followed for its
derivation. Therefore, the existence of a preferred configuration
corresponding to the one at the time of pinning, its stabilization by
the pinning disorder, and the resulting dynamical coexistence between
configurations that are similar to the reference one and
configurations that are significantly different might be generic
qualitative features of the PP systems, not necessarily tied to any
particular underlying scenario.
\section{Conclusion}
\label{sec:conclusion}

In the recent years, a significant research effort has been invested
in studies of partly pinned systems prepared from glassforming
liquids. A strong motivation for this comes from a number of
theoretical breakthroughs, suggesting that such studies have the
potential to provide clear insight into pending issues of the physics
of glassy systems. For instance, the theory of point-to-set
correlations shows how PP systems could be used to detect the growth
of otherwise elusive forms of ``amorphous order'' when approaching the
glass transition. \cite{BouBir04JCP,MonSem06JSP_2,%
  MezMon06JSP,MonSem06JSP,FraMon07JPA,ZarFra10JSM,FraSem11book}
Another remarkable example is the prediction, within the RFOT theory,
of an entropy-vanishing ideal liquid-glass transition in homogeneous
PP systems that could be studied fully in equilibrium, at variance
with the putative one in the bulk. \cite{CamBir12PNAS,CamBir13JCP,%
  KobBer13PRL}

But, when a glassforming liquid is turned into a PP system, the
pinning process does not only couple to its glassy
properties. Physical phenomena already present in the PP systems
deriving from normal fluid states and pertaining to the physics of
adsorption on random substrates also come into play and blend with
those related to glassiness.

The aim of the present work was to discuss this phenomena and to
outline a reference framework allowing one to take them into account
in studies of glassy systems.  Indeed, we believe, and we have
provided arguments supporting this idea, that approaches
systematically contrasting the behaviors of the glassy and normal PP
systems should lead to sharper characterizations of the physics
related to the glass transition than approaches only concentrating on
the former.

This will require a better qualitative and quantitative knowledge of
the properties of the normal PP systems than presently
available. Indeed, these systems have been rather neglected in
comparison to their glassy counterparts. The development of accurate
integral equation theories based on closures of the Ornstein-Zernike
equations \eqref{eq:disOZ} and \eqref{eq:disOZhet} would be
particularly interesting, because they could then be used to pinpoint
the regime in which one needs to invoke new physics to explain the
actual behavior of a dense PP fluid system. This will be the goal
pursued in future work.

\begin{acknowledgments}
  Part of this work has been developed during stays at the university
  of Konstanz and at the DLR in Cologne. We thank M.~Fuchs,
  Th.~Voigtmann, and their research groups for their kind hospitality
  and useful discussions, and the DFG research unit FOR 1394
  ``Nonlinear response to probe vitrification'' for financial
  support. Funding for this project was provided by a grant from
  R\'egion Rh\^one-Alpes.
\end{acknowledgments}

\appendix

\section{Derivation of Equation \eqref{eq:asymhet}} \label{appB}

In this Appendix, a simple derivation of Eq.~\eqref{eq:asymhet} is
outlined, based on standard arguments from the study of the asymptotic
behavior of the correlation functions in simple
fluids. \cite{Att02book}

Using the decomposition
\begin{multline}\label{eq:decomp}
  \rho c \otimes h(|\mathbf{x}-\mathbf{y}|) = \rho \int d\mathbf{u}
  c(|\mathbf{x}-\mathbf{u}|) \chi_{\mathcal{R}}(\mathbf{u})
  h(|\mathbf{u}-\mathbf{y}|) \\ + \rho \int d\mathbf{u}
  c(|\mathbf{x}-\mathbf{u}|) \chi_{\overline{\mathcal{R}}}(\mathbf{u})
  h(|\mathbf{u}-\mathbf{y}|),
\end{multline}
Eq.~\eqref{eq:disOZhet} can be straightforwardly rewritten as
\begin{equation}\begin{split}
    h_\text{dis}(\mathbf{x},\mathbf{y}) = &
    c_\text{dis}(\mathbf{x},\mathbf{y}) \\ & + \rho \int d\mathbf{u}
    c(|\mathbf{x}-\mathbf{u}|)
    \chi_{\overline{\mathcal{R}}}(\mathbf{u})
    h(|\mathbf{u}-\mathbf{y}|) \\ & + \rho \int d\mathbf{u}
    c_\text{dis}(\mathbf{x},\mathbf{u}) \chi_{\mathcal{R}}(\mathbf{u})
    h(|\mathbf{u}-\mathbf{y}|) \\ & + \rho \int d\mathbf{u}
    c(|\mathbf{x}-\mathbf{u}|) \chi_{\mathcal{R}}(\mathbf{u})
    h_\text{dis}(\mathbf{u},\mathbf{y}) \\ & - \rho \int d\mathbf{u}
    c_\text{dis}(\mathbf{x},\mathbf{u}) \chi_{\mathcal{R}}(\mathbf{u})
    h_\text{dis}(\mathbf{u},\mathbf{y}).
\end{split}\end{equation}
For $\mathbf{x}$ deep in the fluid region $\mathcal{R}$, the
topological constraints on the diagrams contributing to $c_\text{dis}$
and $h_\text{dis}$ imply that the terms involving $c_\text{dis}$ can
be neglected. Indeed, the leading diagrams in $c_\text{dis}$ are more
connected in terms of fluid-matrix bonds than those in
$h_\text{dis}$. \cite{Note2} It follows that $c_\text{dis}$ decays
much faster than $h_\text{dis}$ when the distance from the random
substrate increases. Therefore, in this regime, one is left with
\begin{multline}\label{eq:approx1}
  h_\text{dis}(\mathbf{x},\mathbf{y}) \simeq \rho \int d\mathbf{u}
  c(|\mathbf{x}-\mathbf{u}|) \chi_{\overline{\mathcal{R}}}(\mathbf{u})
  h(|\mathbf{u}-\mathbf{y}|) \\ + \rho \int d\mathbf{u}
  c(|\mathbf{x}-\mathbf{u}|) \chi_{\mathcal{R}}(\mathbf{u})
  h_\text{dis}(\mathbf{u},\mathbf{y}).
\end{multline}

One can also use Eq.~\eqref{eq:decomp} in the OZ relation for the
original bulk system, to get
\begin{multline}
  h(|\mathbf{x}-\mathbf{y}|) = c(|\mathbf{x}-\mathbf{y}|) + \rho \int
  d\mathbf{u} c(|\mathbf{x}-\mathbf{u}|)
  \chi_{\mathcal{R}}(\mathbf{u}) h(|\mathbf{u}-\mathbf{y}|) \\ + \rho
  \int d\mathbf{u} c(|\mathbf{x}-\mathbf{u}|)
  \chi_{\overline{\mathcal{R}}}(\mathbf{u})
  h(|\mathbf{u}-\mathbf{y}|),
\end{multline}
For $\mathbf{x}$ deep in the fluid region $\mathcal{R}$, the last term
is very small because of the factor $c(|\mathbf{x}-\mathbf{u}|)$ in
the integrand, with $\mathbf{u}$ constrained to be in
$\overline{\mathcal{R}}$. So, one might approximate
\begin{equation}\label{eq:approx2}
  h(|\mathbf{x}-\mathbf{y}|) \simeq c(|\mathbf{x}-\mathbf{y}|) + \rho
  \int d\mathbf{u} c(|\mathbf{x}-\mathbf{u}|)
  \chi_{\mathcal{R}}(\mathbf{u}) h(|\mathbf{u}-\mathbf{y}|).
\end{equation}

Finally, Eqs. \eqref{eq:approx1} and \eqref{eq:approx2} can be
recursively solved for $h_\text{dis}(\mathbf{x},\mathbf{y})$ and
$h(|\mathbf{x}-\mathbf{y}|)$, respectively. Comparing the two
resulting infinite sums, one immediately recognizes
\begin{equation}
  h_\text{dis}(\mathbf{x},\mathbf{y}) \simeq \rho \int d\mathbf{u}
  h(|\mathbf{x}-\mathbf{u}|) \chi_{\overline{\mathcal{R}}}(\mathbf{u})
  h(|\mathbf{u}-\mathbf{y}|).
\end{equation}
Since $h_\text{dis}(\mathbf{x},\mathbf{y})$ is invariant by exchange
of its variables, it is obvious that this equation remains valid if it
is $\mathbf{y}$ and not $\mathbf{x}$ that lies deep in the fluid
region.


\begin{thebibliography}{93}%
\makeatletter
\providecommand \@ifxundefined [1]{%
 \@ifx{#1\undefined}
}%
\providecommand \@ifnum [1]{%
 \ifnum #1\expandafter \@firstoftwo
 \else \expandafter \@secondoftwo
 \fi
}%
\providecommand \@ifx [1]{%
 \ifx #1\expandafter \@firstoftwo
 \else \expandafter \@secondoftwo
 \fi
}%
\providecommand \natexlab [1]{#1}%
\providecommand \enquote  [1]{``#1''}%
\providecommand \bibnamefont  [1]{#1}%
\providecommand \bibfnamefont [1]{#1}%
\providecommand \citenamefont [1]{#1}%
\providecommand \href@noop [0]{\@secondoftwo}%
\providecommand \href [0]{\begingroup \@sanitize@url \@href}%
\providecommand \@href[1]{\@@startlink{#1}\@@href}%
\providecommand \@@href[1]{\endgroup#1\@@endlink}%
\providecommand \@sanitize@url [0]{\catcode `\\12\catcode `\$12\catcode
  `\&12\catcode `\#12\catcode `\^12\catcode `\_12\catcode `\%12\relax}%
\providecommand \@@startlink[1]{}%
\providecommand \@@endlink[0]{}%
\providecommand \url  [0]{\begingroup\@sanitize@url \@url }%
\providecommand \@url [1]{\endgroup\@href {#1}{\urlprefix }}%
\providecommand \urlprefix  [0]{URL }%
\providecommand \Eprint [0]{\href }%
\providecommand \doibase [0]{http://dx.doi.org/}%
\providecommand \selectlanguage [0]{\@gobble}%
\providecommand \bibinfo  [0]{\@secondoftwo}%
\providecommand \bibfield  [0]{\@secondoftwo}%
\providecommand \translation [1]{[#1]}%
\providecommand \BibitemOpen [0]{}%
\providecommand \bibitemStop [0]{}%
\providecommand \bibitemNoStop [0]{.\EOS\space}%
\providecommand \EOS [0]{\spacefactor3000\relax}%
\providecommand \BibitemShut  [1]{\csname bibitem#1\endcsname}%
\let\auto@bib@innerbib\@empty
\bibitem [{\citenamefont {Lifshits}, \citenamefont {Gredeskul},\ and\
  \citenamefont {Pastur}(1988)}]{LifGrePasbook}%
  \BibitemOpen
  \bibfield  {author} {\bibinfo {author} {\bibfnamefont {I.~M.}\ \bibnamefont
  {Lifshits}}, \bibinfo {author} {\bibfnamefont {S.~A.}\ \bibnamefont
  {Gredeskul}}, \ and\ \bibinfo {author} {\bibfnamefont {L.~A.}\ \bibnamefont
  {Pastur}},\ }\href@noop {} {\emph {\bibinfo {title} {Introduction to the
  theory of disordered systems}}}\ (\bibinfo  {publisher} {Wiley},\ \bibinfo
  {address} {New York},\ \bibinfo {year} {1988})\BibitemShut {NoStop}%
\bibitem [{\citenamefont {Scheidler}, \citenamefont {Kob},\ and\ \citenamefont
  {Binder}(2004)}]{SchKobBin04JPCB}%
  \BibitemOpen
  \bibfield  {author} {\bibinfo {author} {\bibfnamefont {P.}~\bibnamefont
  {Scheidler}}, \bibinfo {author} {\bibfnamefont {W.}~\bibnamefont {Kob}}, \
  and\ \bibinfo {author} {\bibfnamefont {K.}~\bibnamefont {Binder}},\ }\href
  {\doibase 10.1021/jp036593s} {\bibfield  {journal} {\bibinfo  {journal} {J.
  Phys. Chem. B}\ }\textbf {\bibinfo {volume} {108}},\ \bibinfo {pages} {6673}
  (\bibinfo {year} {2004})}\BibitemShut {NoStop}%
\bibitem [{\citenamefont {Krakoviack}(2010)}]{Kra10PRE}%
  \BibitemOpen
  \bibfield  {author} {\bibinfo {author} {\bibfnamefont {V.}~\bibnamefont
  {Krakoviack}},\ }\href {\doibase 10.1103/PhysRevE.82.061501} {\bibfield
  {journal} {\bibinfo  {journal} {Phys. Rev. E}\ }\textbf {\bibinfo {volume}
  {82}},\ \bibinfo {pages} {061501} (\bibinfo {year} {2010})}\BibitemShut
  {NoStop}%
\bibitem [{\citenamefont {Cavagna}, \citenamefont {Grigera},\ and\
  \citenamefont {Verrocchio}(2010)}]{CavGriVer10JSMTE}%
  \BibitemOpen
  \bibfield  {author} {\bibinfo {author} {\bibfnamefont {A.}~\bibnamefont
  {Cavagna}}, \bibinfo {author} {\bibfnamefont {T.~S.}\ \bibnamefont
  {Grigera}}, \ and\ \bibinfo {author} {\bibfnamefont {P.}~\bibnamefont
  {Verrocchio}},\ }\href {\doibase 10.1088/1742-5468/2010/10/P10001} {\bibfield
   {journal} {\bibinfo  {journal} {J. Stat. Mech.}\ ,\ \bibinfo {pages}
  {P10001}} (\bibinfo {year} {2010})}\BibitemShut {NoStop}%
\bibitem [{\citenamefont {Franz}\ and\ \citenamefont
  {Semerjian}(2011)}]{FraSem11book}%
  \BibitemOpen
  \bibfield  {author} {\bibinfo {author} {\bibfnamefont {S.}~\bibnamefont
  {Franz}}\ and\ \bibinfo {author} {\bibfnamefont {G.}~\bibnamefont
  {Semerjian}},\ }\enquote {\bibinfo {title} {Dynamical heterogeneities in
  glasses, colloids and granular materials},}\ \ (\bibinfo  {publisher} {Oxford
  University Press},\ \bibinfo {address} {Oxford},\ \bibinfo {year} {2011})\
  Chap.\ \bibinfo {chapter} {Analytical approaches to time and length scales in
  models of glasses}\BibitemShut {NoStop}%
\bibitem [{\citenamefont {Viramontes-Gamboa}, \citenamefont {Arauz-Lara},\ and\
  \citenamefont {Medina-Noyola}(1995)}]{VirAraMed95PRL}%
  \BibitemOpen
  \bibfield  {author} {\bibinfo {author} {\bibfnamefont {G.}~\bibnamefont
  {Viramontes-Gamboa}}, \bibinfo {author} {\bibfnamefont {J.~L.}\ \bibnamefont
  {Arauz-Lara}}, \ and\ \bibinfo {author} {\bibfnamefont {M.}~\bibnamefont
  {Medina-Noyola}},\ }\href {\doibase 10.1103/PhysRevLett.75.759} {\bibfield
  {journal} {\bibinfo  {journal} {Phys. Rev. Lett.}\ }\textbf {\bibinfo
  {volume} {75}},\ \bibinfo {pages} {759} (\bibinfo {year} {1995})}\BibitemShut
  {NoStop}%
\bibitem [{\citenamefont {Viramontes-Gamboa}, \citenamefont {Medina-Noyola},\
  and\ \citenamefont {Arauz-Lara}(1995)}]{VirMedAra95PRE}%
  \BibitemOpen
  \bibfield  {author} {\bibinfo {author} {\bibfnamefont {G.}~\bibnamefont
  {Viramontes-Gamboa}}, \bibinfo {author} {\bibfnamefont {M.}~\bibnamefont
  {Medina-Noyola}}, \ and\ \bibinfo {author} {\bibfnamefont {J.~L.}\
  \bibnamefont {Arauz-Lara}},\ }\href {\doibase 10.1103/PhysRevE.52.4035}
  {\bibfield  {journal} {\bibinfo  {journal} {Phys. Rev. E}\ }\textbf {\bibinfo
  {volume} {52}},\ \bibinfo {pages} {4035} (\bibinfo {year}
  {1995})}\BibitemShut {NoStop}%
\bibitem [{\citenamefont {Ch{\'a}vez-Rojo}, \citenamefont
  {Ju{\'a}rez-Maldonado},\ and\ \citenamefont
  {Medina-Noyola}(2008)}]{ChaJuaMed08PRE}%
  \BibitemOpen
  \bibfield  {author} {\bibinfo {author} {\bibfnamefont {M.~A.}\ \bibnamefont
  {Ch{\'a}vez-Rojo}}, \bibinfo {author} {\bibfnamefont {R.}~\bibnamefont
  {Ju{\'a}rez-Maldonado}}, \ and\ \bibinfo {author} {\bibfnamefont
  {M.}~\bibnamefont {Medina-Noyola}},\ }\href {\doibase
  10.1103/PhysRevE.77.040401} {\bibfield  {journal} {\bibinfo  {journal} {Phys.
  Rev. E}\ }\textbf {\bibinfo {volume} {77}},\ \bibinfo {pages} {040401}
  (\bibinfo {year} {2008})}\BibitemShut {NoStop}%
\bibitem [{\citenamefont {Scheidler}\ \emph {et~al.}(2002)\citenamefont
  {Scheidler}, \citenamefont {Kob}, \citenamefont {Binder},\ and\ \citenamefont
  {Parisi}}]{SchKobBinPar02PMB}%
  \BibitemOpen
  \bibfield  {author} {\bibinfo {author} {\bibfnamefont {P.}~\bibnamefont
  {Scheidler}}, \bibinfo {author} {\bibfnamefont {W.}~\bibnamefont {Kob}},
  \bibinfo {author} {\bibfnamefont {K.}~\bibnamefont {Binder}}, \ and\ \bibinfo
  {author} {\bibfnamefont {G.}~\bibnamefont {Parisi}},\ }\href {\doibase
  10.1080/13642810208221307} {\bibfield  {journal} {\bibinfo  {journal} {Phil.
  Mag. B}\ }\textbf {\bibinfo {volume} {82}},\ \bibinfo {pages} {283} (\bibinfo
  {year} {2002})}\BibitemShut {NoStop}%
\bibitem [{\citenamefont {Scheidler}, \citenamefont {Kob},\ and\ \citenamefont
  {Binder}(2002)}]{SchKobBin02EL}%
  \BibitemOpen
  \bibfield  {author} {\bibinfo {author} {\bibfnamefont {P.}~\bibnamefont
  {Scheidler}}, \bibinfo {author} {\bibfnamefont {W.}~\bibnamefont {Kob}}, \
  and\ \bibinfo {author} {\bibfnamefont {K.}~\bibnamefont {Binder}},\ }\href
  {\doibase 10.1209/epl/i2002-00182-9} {\bibfield  {journal} {\bibinfo
  {journal} {Europhys. Lett.}\ }\textbf {\bibinfo {volume} {59}},\ \bibinfo
  {pages} {701} (\bibinfo {year} {2002})}\BibitemShut {NoStop}%
\bibitem [{\citenamefont {Scheidler}, \citenamefont {Kob},\ and\ \citenamefont
  {Binder}(2003)}]{SchKobBin03EPJE}%
  \BibitemOpen
  \bibfield  {author} {\bibinfo {author} {\bibfnamefont {P.}~\bibnamefont
  {Scheidler}}, \bibinfo {author} {\bibfnamefont {W.}~\bibnamefont {Kob}}, \
  and\ \bibinfo {author} {\bibfnamefont {K.}~\bibnamefont {Binder}},\ }\href
  {\doibase 10.1140/epje/i2003-10041-7} {\bibfield  {journal} {\bibinfo
  {journal} {Eur. Phys. J. E}\ }\textbf {\bibinfo {volume} {12}},\ \bibinfo
  {pages} {5} (\bibinfo {year} {2003})}\BibitemShut {NoStop}%
\bibitem [{\citenamefont {Kim}(2003)}]{Kim03EL}%
  \BibitemOpen
  \bibfield  {author} {\bibinfo {author} {\bibfnamefont {K.}~\bibnamefont
  {Kim}},\ }\href {\doibase 10.1209/epl/i2003-00303-0} {\bibfield  {journal}
  {\bibinfo  {journal} {Europhys. Lett.}\ }\textbf {\bibinfo {volume} {61}},\
  \bibinfo {pages} {790} (\bibinfo {year} {2003})}\BibitemShut {NoStop}%
\bibitem [{\citenamefont {Chang}, \citenamefont {Jagannathan},\ and\
  \citenamefont {Yethiraj}(2004)}]{ChaJagYet04PRE}%
  \BibitemOpen
  \bibfield  {author} {\bibinfo {author} {\bibfnamefont {R.}~\bibnamefont
  {Chang}}, \bibinfo {author} {\bibfnamefont {K.}~\bibnamefont {Jagannathan}},
  \ and\ \bibinfo {author} {\bibfnamefont {A.}~\bibnamefont {Yethiraj}},\
  }\href {\doibase 10.1103/PhysRevE.69.051101} {\bibfield  {journal} {\bibinfo
  {journal} {Phys. Rev. E}\ }\textbf {\bibinfo {volume} {69}},\ \bibinfo
  {pages} {051101} (\bibinfo {year} {2004})}\BibitemShut {NoStop}%
\bibitem [{\citenamefont {Mittal}, \citenamefont {Errington},\ and\
  \citenamefont {Truskett}(2006)}]{MitErrTru06PRE}%
  \BibitemOpen
  \bibfield  {author} {\bibinfo {author} {\bibfnamefont {J.}~\bibnamefont
  {Mittal}}, \bibinfo {author} {\bibfnamefont {J.~R.}\ \bibnamefont
  {Errington}}, \ and\ \bibinfo {author} {\bibfnamefont {T.~M.}\ \bibnamefont
  {Truskett}},\ }\href {\doibase 10.1103/PhysRevE.74.040102} {\bibfield
  {journal} {\bibinfo  {journal} {Phys. Rev. E}\ }\textbf {\bibinfo {volume}
  {74}},\ \bibinfo {pages} {040102} (\bibinfo {year} {2006})}\BibitemShut
  {NoStop}%
\bibitem [{\citenamefont {Kim}, \citenamefont {Miyazaki},\ and\ \citenamefont
  {Saito}(2009)}]{KimMiySai09EL}%
  \BibitemOpen
  \bibfield  {author} {\bibinfo {author} {\bibfnamefont {K.}~\bibnamefont
  {Kim}}, \bibinfo {author} {\bibfnamefont {K.}~\bibnamefont {Miyazaki}}, \
  and\ \bibinfo {author} {\bibfnamefont {S.}~\bibnamefont {Saito}},\ }\href
  {\doibase 10.1209/0295-5075/88/36002} {\bibfield  {journal} {\bibinfo
  {journal} {EPL}\ }\textbf {\bibinfo {volume} {88}},\ \bibinfo {pages} {36002}
  (\bibinfo {year} {2009})}\BibitemShut {NoStop}%
\bibitem [{\citenamefont {Kim}, \citenamefont {Miyazaki},\ and\ \citenamefont
  {Saito}(2010)}]{KimMiySai10EPJST}%
  \BibitemOpen
  \bibfield  {author} {\bibinfo {author} {\bibfnamefont {K.}~\bibnamefont
  {Kim}}, \bibinfo {author} {\bibfnamefont {K.}~\bibnamefont {Miyazaki}}, \
  and\ \bibinfo {author} {\bibfnamefont {S.}~\bibnamefont {Saito}},\ }\href
  {\doibase 10.1140/epjst/e2010-01315-y} {\bibfield  {journal} {\bibinfo
  {journal} {Eur. Phys. J. Special Topics}\ }\textbf {\bibinfo {volume}
  {189}},\ \bibinfo {pages} {135} (\bibinfo {year} {2010})}\BibitemShut
  {NoStop}%
\bibitem [{\citenamefont {Kim}, \citenamefont {Miyazaki},\ and\ \citenamefont
  {Saito}(2011)}]{KimMiySai11JPCM}%
  \BibitemOpen
  \bibfield  {author} {\bibinfo {author} {\bibfnamefont {K.}~\bibnamefont
  {Kim}}, \bibinfo {author} {\bibfnamefont {K.}~\bibnamefont {Miyazaki}}, \
  and\ \bibinfo {author} {\bibfnamefont {S.}~\bibnamefont {Saito}},\ }\href
  {\doibase 10.1088/0953-8984/23/23/234123} {\bibfield  {journal} {\bibinfo
  {journal} {J. Phys.: Condens. Matter}\ }\textbf {\bibinfo {volume} {23}},\
  \bibinfo {pages} {234123} (\bibinfo {year} {2011})}\BibitemShut {NoStop}%
\bibitem [{\citenamefont {Fenz}\ \emph {et~al.}(2009)\citenamefont {Fenz},
  \citenamefont {Mryglod}, \citenamefont {Prytula},\ and\ \citenamefont
  {Folk}}]{FenMryPryFol09PRE}%
  \BibitemOpen
  \bibfield  {author} {\bibinfo {author} {\bibfnamefont {W.}~\bibnamefont
  {Fenz}}, \bibinfo {author} {\bibfnamefont {I.~M.}\ \bibnamefont {Mryglod}},
  \bibinfo {author} {\bibfnamefont {O.}~\bibnamefont {Prytula}}, \ and\
  \bibinfo {author} {\bibfnamefont {R.}~\bibnamefont {Folk}},\ }\href {\doibase
  10.1103/PhysRevE.80.021202} {\bibfield  {journal} {\bibinfo  {journal} {Phys.
  Rev. E}\ }\textbf {\bibinfo {volume} {80}},\ \bibinfo {pages} {021202}
  (\bibinfo {year} {2009})}\BibitemShut {NoStop}%
\bibitem [{\citenamefont {Karmakar}, \citenamefont {Lerner},\ and\
  \citenamefont {Procaccia}(2012)}]{KarLerPro12PA}%
  \BibitemOpen
  \bibfield  {author} {\bibinfo {author} {\bibfnamefont {S.}~\bibnamefont
  {Karmakar}}, \bibinfo {author} {\bibfnamefont {E.}~\bibnamefont {Lerner}}, \
  and\ \bibinfo {author} {\bibfnamefont {I.}~\bibnamefont {Procaccia}},\ }\href
  {\doibase 10.1016/j.physa.2011.11.020} {\bibfield  {journal} {\bibinfo
  {journal} {Physica A}\ }\textbf {\bibinfo {volume} {391}},\ \bibinfo {pages}
  {1001} (\bibinfo {year} {2012})}\BibitemShut {NoStop}%
\bibitem [{\citenamefont {Kob}, \citenamefont {Rold{\'a}n-Vargas},\ and\
  \citenamefont {Berthier}(2012{\natexlab{a}})}]{KobRolBer12NatPhys}%
  \BibitemOpen
  \bibfield  {author} {\bibinfo {author} {\bibfnamefont {W.}~\bibnamefont
  {Kob}}, \bibinfo {author} {\bibfnamefont {S.}~\bibnamefont
  {Rold{\'a}n-Vargas}}, \ and\ \bibinfo {author} {\bibfnamefont
  {L.}~\bibnamefont {Berthier}},\ }\href {\doibase 10.1038/nphys2133}
  {\bibfield  {journal} {\bibinfo  {journal} {Nature Phys.}\ }\textbf {\bibinfo
  {volume} {8}},\ \bibinfo {pages} {164} (\bibinfo {year}
  {2012}{\natexlab{a}})}\BibitemShut {NoStop}%
\bibitem [{\citenamefont {Kob}, \citenamefont {Rold{\'a}n-Vargas},\ and\
  \citenamefont {Berthier}(2012{\natexlab{b}})}]{KobRolBer12PhysProc}%
  \BibitemOpen
  \bibfield  {author} {\bibinfo {author} {\bibfnamefont {W.}~\bibnamefont
  {Kob}}, \bibinfo {author} {\bibfnamefont {S.}~\bibnamefont
  {Rold{\'a}n-Vargas}}, \ and\ \bibinfo {author} {\bibfnamefont
  {L.}~\bibnamefont {Berthier}},\ }\href {\doibase 10.1016/j.phpro.2012.05.012}
  {\bibfield  {journal} {\bibinfo  {journal} {Phys. Procedia}\ }\textbf
  {\bibinfo {volume} {34}},\ \bibinfo {pages} {70–79} (\bibinfo {year}
  {2012}{\natexlab{b}})}\BibitemShut {NoStop}%
\bibitem [{\citenamefont {Klameth}\ and\ \citenamefont
  {Vogel}(2013)}]{KlaVog13JCP}%
  \BibitemOpen
  \bibfield  {author} {\bibinfo {author} {\bibfnamefont {F.}~\bibnamefont
  {Klameth}}\ and\ \bibinfo {author} {\bibfnamefont {M.}~\bibnamefont
  {Vogel}},\ }\href {\doibase 10.1063/1.4798217} {\bibfield  {journal}
  {\bibinfo  {journal} {J. Chem. Phys.}\ }\textbf {\bibinfo {volume} {138}},\
  \bibinfo {pages} {134503} (\bibinfo {year} {2013})}\BibitemShut {NoStop}%
\bibitem [{\citenamefont {Klameth}, \citenamefont {Henritzi},\ and\
  \citenamefont {Vogel}(2014)}]{KlaHenVog14JCP}%
  \BibitemOpen
  \bibfield  {author} {\bibinfo {author} {\bibfnamefont {F.}~\bibnamefont
  {Klameth}}, \bibinfo {author} {\bibfnamefont {P.}~\bibnamefont {Henritzi}}, \
  and\ \bibinfo {author} {\bibfnamefont {M.}~\bibnamefont {Vogel}},\
  }\href@noop {} {\bibfield  {journal} {\bibinfo  {journal} {J. Chem. Phys.}\
  }\textbf {\bibinfo {volume} {140}},\ \bibinfo {pages} {144501} (\bibinfo
  {year} {2014})}\BibitemShut {NoStop}%
\bibitem [{\citenamefont {Bouchaud}\ and\ \citenamefont
  {Biroli}(2004)}]{BouBir04JCP}%
  \BibitemOpen
  \bibfield  {author} {\bibinfo {author} {\bibfnamefont {J.-P.}\ \bibnamefont
  {Bouchaud}}\ and\ \bibinfo {author} {\bibfnamefont {G.}~\bibnamefont
  {Biroli}},\ }\href {\doibase 10.1063/1.1796231} {\bibfield  {journal}
  {\bibinfo  {journal} {J. Chem. Phys.}\ }\textbf {\bibinfo {volume} {121}},\
  \bibinfo {pages} {7347} (\bibinfo {year} {2004})}\BibitemShut {NoStop}%
\bibitem [{\citenamefont {Montanari}\ and\ \citenamefont
  {Semerjian}(2006{\natexlab{a}})}]{MonSem06JSP_2}%
  \BibitemOpen
  \bibfield  {author} {\bibinfo {author} {\bibfnamefont {A.}~\bibnamefont
  {Montanari}}\ and\ \bibinfo {author} {\bibfnamefont {G.}~\bibnamefont
  {Semerjian}},\ }\href {\doibase 10.1007/s10955-006-9103-1} {\bibfield
  {journal} {\bibinfo  {journal} {J. Stat. Phys.}\ }\textbf {\bibinfo {volume}
  {124}},\ \bibinfo {pages} {103} (\bibinfo {year}
  {2006}{\natexlab{a}})}\BibitemShut {NoStop}%
\bibitem [{\citenamefont {M{\'e}zard}\ and\ \citenamefont
  {Montanari}(2006)}]{MezMon06JSP}%
  \BibitemOpen
  \bibfield  {author} {\bibinfo {author} {\bibfnamefont {M.}~\bibnamefont
  {M{\'e}zard}}\ and\ \bibinfo {author} {\bibfnamefont {A.}~\bibnamefont
  {Montanari}},\ }\href {\doibase 10.1007/s10955-006-9162-3} {\bibfield
  {journal} {\bibinfo  {journal} {J. Stat. Phys.}\ }\textbf {\bibinfo {volume}
  {124}},\ \bibinfo {pages} {1317} (\bibinfo {year} {2006})}\BibitemShut
  {NoStop}%
\bibitem [{\citenamefont {Montanari}\ and\ \citenamefont
  {Semerjian}(2006{\natexlab{b}})}]{MonSem06JSP}%
  \BibitemOpen
  \bibfield  {author} {\bibinfo {author} {\bibfnamefont {A.}~\bibnamefont
  {Montanari}}\ and\ \bibinfo {author} {\bibfnamefont {G.}~\bibnamefont
  {Semerjian}},\ }\href {\doibase 10.1007/s10955-006-9175-y} {\bibfield
  {journal} {\bibinfo  {journal} {J. Stat. Phys.}\ }\textbf {\bibinfo {volume}
  {125}},\ \bibinfo {pages} {23} (\bibinfo {year}
  {2006}{\natexlab{b}})}\BibitemShut {NoStop}%
\bibitem [{\citenamefont {Franz}\ and\ \citenamefont
  {Montanari}(2007)}]{FraMon07JPA}%
  \BibitemOpen
  \bibfield  {author} {\bibinfo {author} {\bibfnamefont {S.}~\bibnamefont
  {Franz}}\ and\ \bibinfo {author} {\bibfnamefont {A.}~\bibnamefont
  {Montanari}},\ }\href {\doibase 10.1088/1751-8113/40/11/F01} {\bibfield
  {journal} {\bibinfo  {journal} {J. Phys. A: Math. Theor.}\ }\textbf {\bibinfo
  {volume} {40}},\ \bibinfo {pages} {F251} (\bibinfo {year}
  {2007})}\BibitemShut {NoStop}%
\bibitem [{\citenamefont {Zarinelli}\ and\ \citenamefont
  {Franz}(2010)}]{ZarFra10JSM}%
  \BibitemOpen
  \bibfield  {author} {\bibinfo {author} {\bibfnamefont {E.}~\bibnamefont
  {Zarinelli}}\ and\ \bibinfo {author} {\bibfnamefont {S.}~\bibnamefont
  {Franz}},\ }\href@noop {} {\bibfield  {journal} {\bibinfo  {journal} {J.
  Stat. Mech.}\ ,\ \bibinfo {pages} {P04008}} (\bibinfo {year}
  {2010})}\BibitemShut {NoStop}%
\bibitem [{\citenamefont {Jack}\ and\ \citenamefont
  {Garrahan}(2005)}]{JacGar05JCP}%
  \BibitemOpen
  \bibfield  {author} {\bibinfo {author} {\bibfnamefont {R.~L.}\ \bibnamefont
  {Jack}}\ and\ \bibinfo {author} {\bibfnamefont {J.~P.}\ \bibnamefont
  {Garrahan}},\ }\href {\doibase 10.1063/1.2075067} {\bibfield  {journal}
  {\bibinfo  {journal} {J. Chem. Phys.}\ }\textbf {\bibinfo {volume} {123}},\
  \bibinfo {pages} {164508} (\bibinfo {year} {2005})}\BibitemShut {NoStop}%
\bibitem [{\citenamefont {Cavagna}, \citenamefont {Grigera},\ and\
  \citenamefont {Verrocchio}(2007)}]{CavGriVer07PRL}%
  \BibitemOpen
  \bibfield  {author} {\bibinfo {author} {\bibfnamefont {A.}~\bibnamefont
  {Cavagna}}, \bibinfo {author} {\bibfnamefont {T.~S.}\ \bibnamefont
  {Grigera}}, \ and\ \bibinfo {author} {\bibfnamefont {P.}~\bibnamefont
  {Verrocchio}},\ }\href {\doibase 10.1103/PhysRevLett.98.187801} {\bibfield
  {journal} {\bibinfo  {journal} {Phys. Rev. Lett.}\ }\textbf {\bibinfo
  {volume} {98}},\ \bibinfo {pages} {187801} (\bibinfo {year}
  {2007})}\BibitemShut {NoStop}%
\bibitem [{\citenamefont {Biroli}\ \emph {et~al.}(2008)\citenamefont {Biroli},
  \citenamefont {Bouchaud}, \citenamefont {Cavagna}, \citenamefont {Grigera},\
  and\ \citenamefont {Verrocchio}}]{BirBouCavGriVer08NatPhys}%
  \BibitemOpen
  \bibfield  {author} {\bibinfo {author} {\bibfnamefont {G.}~\bibnamefont
  {Biroli}}, \bibinfo {author} {\bibfnamefont {J.-P.}\ \bibnamefont
  {Bouchaud}}, \bibinfo {author} {\bibfnamefont {A.}~\bibnamefont {Cavagna}},
  \bibinfo {author} {\bibfnamefont {T.~S.}\ \bibnamefont {Grigera}}, \ and\
  \bibinfo {author} {\bibfnamefont {P.}~\bibnamefont {Verrocchio}},\ }\href
  {\doibase 10.1038/nphys1050} {\bibfield  {journal} {\bibinfo  {journal}
  {Nature Phys.}\ }\textbf {\bibinfo {volume} {4}},\ \bibinfo {pages} {771}
  (\bibinfo {year} {2008})}\BibitemShut {NoStop}%
\bibitem [{\citenamefont {Cavagna}, \citenamefont {Grigera},\ and\
  \citenamefont {Verrocchio}(2012)}]{CavGriVer12JCP}%
  \BibitemOpen
  \bibfield  {author} {\bibinfo {author} {\bibfnamefont {A.}~\bibnamefont
  {Cavagna}}, \bibinfo {author} {\bibfnamefont {T.~S.}\ \bibnamefont
  {Grigera}}, \ and\ \bibinfo {author} {\bibfnamefont {P.}~\bibnamefont
  {Verrocchio}},\ }\href {\doibase 10.1063/1.4720477} {\bibfield  {journal}
  {\bibinfo  {journal} {J. Chem. Phys.}\ }\textbf {\bibinfo {volume} {136}},\
  \bibinfo {pages} {204502} (\bibinfo {year} {2012})}\BibitemShut {NoStop}%
\bibitem [{\citenamefont {Gradenigo}\ \emph {et~al.}(2013)\citenamefont
  {Gradenigo}, \citenamefont {Trozzo}, \citenamefont {Cavagna}, \citenamefont
  {Grigera},\ and\ \citenamefont {Verrocchio}}]{GraTroCavGriVer13JCP}%
  \BibitemOpen
  \bibfield  {author} {\bibinfo {author} {\bibfnamefont {G.}~\bibnamefont
  {Gradenigo}}, \bibinfo {author} {\bibfnamefont {R.}~\bibnamefont {Trozzo}},
  \bibinfo {author} {\bibfnamefont {A.}~\bibnamefont {Cavagna}}, \bibinfo
  {author} {\bibfnamefont {T.~S.}\ \bibnamefont {Grigera}}, \ and\ \bibinfo
  {author} {\bibfnamefont {P.}~\bibnamefont {Verrocchio}},\ }\href@noop {}
  {\bibfield  {journal} {\bibinfo  {journal} {J. Chem. Phys.}\ }\textbf
  {\bibinfo {volume} {138}},\ \bibinfo {pages} {12{A}509} (\bibinfo {year}
  {2013})}\BibitemShut {NoStop}%
\bibitem [{\citenamefont {Sausset}\ and\ \citenamefont
  {Levine}(2011)}]{SauLev11PRL}%
  \BibitemOpen
  \bibfield  {author} {\bibinfo {author} {\bibfnamefont {F.}~\bibnamefont
  {Sausset}}\ and\ \bibinfo {author} {\bibfnamefont {D.}~\bibnamefont
  {Levine}},\ }\href@noop {} {\bibfield  {journal} {\bibinfo  {journal} {Phys.
  Rev. Lett.}\ }\textbf {\bibinfo {volume} {107}},\ \bibinfo {pages} {045501}
  (\bibinfo {year} {2011})}\BibitemShut {NoStop}%
\bibitem [{\citenamefont {Hocky}, \citenamefont {Markland},\ and\ \citenamefont
  {Reichman}(2012)}]{HocMarRei12PRL}%
  \BibitemOpen
  \bibfield  {author} {\bibinfo {author} {\bibfnamefont {G.~M.}\ \bibnamefont
  {Hocky}}, \bibinfo {author} {\bibfnamefont {T.~E.}\ \bibnamefont {Markland}},
  \ and\ \bibinfo {author} {\bibfnamefont {D.~R.}\ \bibnamefont {Reichman}},\
  }\href {\doibase 10.1103/PhysRevLett.108.225506} {\bibfield  {journal}
  {\bibinfo  {journal} {Phys. Rev. Lett.}\ }\textbf {\bibinfo {volume} {108}},\
  \bibinfo {pages} {225506} (\bibinfo {year} {2012})}\BibitemShut {NoStop}%
\bibitem [{\citenamefont {Berthier}\ and\ \citenamefont
  {Kob}(2012)}]{BerKob12PRE}%
  \BibitemOpen
  \bibfield  {author} {\bibinfo {author} {\bibfnamefont {L.}~\bibnamefont
  {Berthier}}\ and\ \bibinfo {author} {\bibfnamefont {W.}~\bibnamefont {Kob}},\
  }\href {\doibase 10.1103/PhysRevE.85.011102} {\bibfield  {journal} {\bibinfo
  {journal} {Phys. Rev. E}\ }\textbf {\bibinfo {volume} {85}},\ \bibinfo
  {pages} {011102} (\bibinfo {year} {2012})}\BibitemShut {NoStop}%
\bibitem [{\citenamefont {Charbonneau}, \citenamefont {Charbonneau},\ and\
  \citenamefont {Tarjus}(2012)}]{ChaChaTar12PRL}%
  \BibitemOpen
  \bibfield  {author} {\bibinfo {author} {\bibfnamefont {B.}~\bibnamefont
  {Charbonneau}}, \bibinfo {author} {\bibfnamefont {P.}~\bibnamefont
  {Charbonneau}}, \ and\ \bibinfo {author} {\bibfnamefont {G.}~\bibnamefont
  {Tarjus}},\ }\href {\doibase 10.1103/PhysRevLett.108.035701} {\bibfield
  {journal} {\bibinfo  {journal} {Phys. Rev. Lett.}\ }\textbf {\bibinfo
  {volume} {108}},\ \bibinfo {pages} {035701} (\bibinfo {year}
  {2012})}\BibitemShut {NoStop}%
\bibitem [{\citenamefont {Charbonneau}, \citenamefont {Charbonneau},\ and\
  \citenamefont {Tarjus}(2013)}]{ChaChaTar13JCP}%
  \BibitemOpen
  \bibfield  {author} {\bibinfo {author} {\bibfnamefont {B.}~\bibnamefont
  {Charbonneau}}, \bibinfo {author} {\bibfnamefont {P.}~\bibnamefont
  {Charbonneau}}, \ and\ \bibinfo {author} {\bibfnamefont {G.}~\bibnamefont
  {Tarjus}},\ }\href@noop {} {\bibfield  {journal} {\bibinfo  {journal} {J.
  Chem. Phys.}\ }\textbf {\bibinfo {volume} {138}},\ \bibinfo {pages}
  {12{A}515} (\bibinfo {year} {2013})}\BibitemShut {NoStop}%
\bibitem [{\citenamefont {Charbonneau}\ and\ \citenamefont
  {Tarjus}(2013)}]{ChaTar13PRE}%
  \BibitemOpen
  \bibfield  {author} {\bibinfo {author} {\bibfnamefont {P.}~\bibnamefont
  {Charbonneau}}\ and\ \bibinfo {author} {\bibfnamefont {G.}~\bibnamefont
  {Tarjus}},\ }\href {\doibase 10.1103/PhysRevE.87.042305} {\bibfield
  {journal} {\bibinfo  {journal} {Phys. Rev. E}\ }\textbf {\bibinfo {volume}
  {87}},\ \bibinfo {pages} {042305} (\bibinfo {year} {2013})}\BibitemShut
  {NoStop}%
\bibitem [{\citenamefont {Biroli}, \citenamefont {Karmakar},\ and\
  \citenamefont {Procaccia}(2013)}]{BirKarPro13PRL}%
  \BibitemOpen
  \bibfield  {author} {\bibinfo {author} {\bibfnamefont {G.}~\bibnamefont
  {Biroli}}, \bibinfo {author} {\bibfnamefont {S.}~\bibnamefont {Karmakar}}, \
  and\ \bibinfo {author} {\bibfnamefont {I.}~\bibnamefont {Procaccia}},\ }\href
  {\doibase 10.1103/PhysRevLett.111.165701} {\bibfield  {journal} {\bibinfo
  {journal} {Phys. Rev. Lett.}\ }\textbf {\bibinfo {volume} {111}},\ \bibinfo
  {pages} {165701} (\bibinfo {year} {2013})}\BibitemShut {NoStop}%
\bibitem [{\citenamefont {Li}, \citenamefont {Xu},\ and\ \citenamefont
  {Sun}(2014)}]{LiXuSun14JCP}%
  \BibitemOpen
  \bibfield  {author} {\bibinfo {author} {\bibfnamefont {Y.-W.}\ \bibnamefont
  {Li}}, \bibinfo {author} {\bibfnamefont {W.-S.}\ \bibnamefont {Xu}}, \ and\
  \bibinfo {author} {\bibfnamefont {Z.-Y.}\ \bibnamefont {Sun}},\ }\href@noop
  {} {\bibfield  {journal} {\bibinfo  {journal} {J. Chem. Phys.}\ }\textbf
  {\bibinfo {volume} {140}},\ \bibinfo {pages} {124502} (\bibinfo {year}
  {2014})}\BibitemShut {NoStop}%
\bibitem [{\citenamefont {Cammarota}\ and\ \citenamefont
  {Biroli}(2012{\natexlab{a}})}]{CamBir12EPL}%
  \BibitemOpen
  \bibfield  {author} {\bibinfo {author} {\bibfnamefont {C.}~\bibnamefont
  {Cammarota}}\ and\ \bibinfo {author} {\bibfnamefont {G.}~\bibnamefont
  {Biroli}},\ }\href@noop {} {\bibfield  {journal} {\bibinfo  {journal} {EPL}\
  }\textbf {\bibinfo {volume} {98}},\ \bibinfo {pages} {16011} (\bibinfo {year}
  {2012}{\natexlab{a}})}\BibitemShut {NoStop}%
\bibitem [{\citenamefont {Cammarota}\ and\ \citenamefont
  {Biroli}(2012{\natexlab{b}})}]{CamBir12PNAS}%
  \BibitemOpen
  \bibfield  {author} {\bibinfo {author} {\bibfnamefont {C.}~\bibnamefont
  {Cammarota}}\ and\ \bibinfo {author} {\bibfnamefont {G.}~\bibnamefont
  {Biroli}},\ }\href@noop {} {\bibfield  {journal} {\bibinfo  {journal} {Proc.
  Natl. Acad. Sci. U.S.A.}\ }\textbf {\bibinfo {volume} {109}},\ \bibinfo
  {pages} {8850} (\bibinfo {year} {2012}{\natexlab{b}})}\BibitemShut {NoStop}%
\bibitem [{\citenamefont {Cammarota}\ and\ \citenamefont
  {Biroli}(2013)}]{CamBir13JCP}%
  \BibitemOpen
  \bibfield  {author} {\bibinfo {author} {\bibfnamefont {C.}~\bibnamefont
  {Cammarota}}\ and\ \bibinfo {author} {\bibfnamefont {G.}~\bibnamefont
  {Biroli}},\ }\href@noop {} {\bibfield  {journal} {\bibinfo  {journal} {J.
  Chem. Phys.}\ }\textbf {\bibinfo {volume} {138}},\ \bibinfo {pages}
  {12{A}547} (\bibinfo {year} {2013})}\BibitemShut {NoStop}%
\bibitem [{\citenamefont {Franz}, \citenamefont {Parisi},\ and\ \citenamefont
  {Ricci-Tersenghi}(2013)}]{FraParRic13JSMTE}%
  \BibitemOpen
  \bibfield  {author} {\bibinfo {author} {\bibfnamefont {S.}~\bibnamefont
  {Franz}}, \bibinfo {author} {\bibfnamefont {G.}~\bibnamefont {Parisi}}, \
  and\ \bibinfo {author} {\bibfnamefont {F.}~\bibnamefont {Ricci-Tersenghi}},\
  }\href@noop {} {\bibfield  {journal} {\bibinfo  {journal} {J. Stat. Mech.}\
  ,\ \bibinfo {pages} {L02001}} (\bibinfo {year} {2013})}\BibitemShut {NoStop}%
\bibitem [{\citenamefont {Cammarota}, \citenamefont {Gradenigo},\ and\
  \citenamefont {Biroli}(2013)}]{CamGraBir13PRL}%
  \BibitemOpen
  \bibfield  {author} {\bibinfo {author} {\bibfnamefont {C.}~\bibnamefont
  {Cammarota}}, \bibinfo {author} {\bibfnamefont {G.}~\bibnamefont
  {Gradenigo}}, \ and\ \bibinfo {author} {\bibfnamefont {G.}~\bibnamefont
  {Biroli}},\ }\href {\doibase 10.1103/PhysRevLett.111.107801} {\bibfield
  {journal} {\bibinfo  {journal} {Phys. Rev. Lett.}\ }\textbf {\bibinfo
  {volume} {111}},\ \bibinfo {pages} {107801} (\bibinfo {year}
  {2013})}\BibitemShut {NoStop}%
\bibitem [{\citenamefont {Krakoviack}(2011)}]{Kra11PRE}%
  \BibitemOpen
  \bibfield  {author} {\bibinfo {author} {\bibfnamefont {V.}~\bibnamefont
  {Krakoviack}},\ }\href {\doibase 10.1103/PhysRevE.84.050501} {\bibfield
  {journal} {\bibinfo  {journal} {Phys. Rev. E}\ }\textbf {\bibinfo {volume}
  {84}},\ \bibinfo {pages} {050501(R)} (\bibinfo {year} {2011})}\BibitemShut
  {NoStop}%
\bibitem [{\citenamefont {Szamel}\ and\ \citenamefont
  {Flenner}(2013)}]{SzaFle13EPL}%
  \BibitemOpen
  \bibfield  {author} {\bibinfo {author} {\bibfnamefont {G.}~\bibnamefont
  {Szamel}}\ and\ \bibinfo {author} {\bibfnamefont {E.}~\bibnamefont
  {Flenner}},\ }\href@noop {} {\bibfield  {journal} {\bibinfo  {journal} {EPL}\
  }\textbf {\bibinfo {volume} {101}},\ \bibinfo {pages} {66005} (\bibinfo
  {year} {2013})}\BibitemShut {NoStop}%
\bibitem [{\citenamefont {Jack}\ and\ \citenamefont
  {Berthier}(2012)}]{JacBer12PRE}%
  \BibitemOpen
  \bibfield  {author} {\bibinfo {author} {\bibfnamefont {R.~L.}\ \bibnamefont
  {Jack}}\ and\ \bibinfo {author} {\bibfnamefont {L.}~\bibnamefont
  {Berthier}},\ }\href@noop {} {\bibfield  {journal} {\bibinfo  {journal}
  {Phys. Rev. E}\ }\textbf {\bibinfo {volume} {85}},\ \bibinfo {pages} {021120}
  (\bibinfo {year} {2012})}\BibitemShut {NoStop}%
\bibitem [{\citenamefont {Kob}\ and\ \citenamefont
  {Berthier}(2013)}]{KobBer13PRL}%
  \BibitemOpen
  \bibfield  {author} {\bibinfo {author} {\bibfnamefont {W.}~\bibnamefont
  {Kob}}\ and\ \bibinfo {author} {\bibfnamefont {L.}~\bibnamefont {Berthier}},\
  }\href {\doibase 10.1103/PhysRevE.85.011102} {\bibfield  {journal} {\bibinfo
  {journal} {Phys. Rev. Lett.}\ }\textbf {\bibinfo {volume} {110}},\ \bibinfo
  {pages} {245702} (\bibinfo {year} {2013})}\BibitemShut {NoStop}%
\bibitem [{\citenamefont {Jack}\ and\ \citenamefont
  {Fullerton}(2013)}]{JacFul13PRE}%
  \BibitemOpen
  \bibfield  {author} {\bibinfo {author} {\bibfnamefont {R.~L.}\ \bibnamefont
  {Jack}}\ and\ \bibinfo {author} {\bibfnamefont {C.~J.}\ \bibnamefont
  {Fullerton}},\ }\href {\doibase 10.1103/PhysRevE.88.042304} {\bibfield
  {journal} {\bibinfo  {journal} {Phys. Rev. E}\ }\textbf {\bibinfo {volume}
  {88}},\ \bibinfo {pages} {042304} (\bibinfo {year} {2013})}\BibitemShut
  {NoStop}%
\bibitem [{\citenamefont {Madden}\ and\ \citenamefont
  {Glandt}(1988)}]{MadGla88JSP}%
  \BibitemOpen
  \bibfield  {author} {\bibinfo {author} {\bibfnamefont {W.~G.}\ \bibnamefont
  {Madden}}\ and\ \bibinfo {author} {\bibfnamefont {E.~D.}\ \bibnamefont
  {Glandt}},\ }\href {\doibase 10.1007/BF01028471} {\bibfield  {journal}
  {\bibinfo  {journal} {J. Stat. Phys.}\ }\textbf {\bibinfo {volume} {51}},\
  \bibinfo {pages} {537} (\bibinfo {year} {1988})}\BibitemShut {NoStop}%
\bibitem [{\citenamefont {Madden}(1992)}]{Mad92JCP}%
  \BibitemOpen
  \bibfield  {author} {\bibinfo {author} {\bibfnamefont {W.~G.}\ \bibnamefont
  {Madden}},\ }\href {\doibase 10.1063/1.462726} {\bibfield  {journal}
  {\bibinfo  {journal} {J. Chem. Phys.}\ }\textbf {\bibinfo {volume} {96}},\
  \bibinfo {pages} {5422} (\bibinfo {year} {1992})}\BibitemShut {NoStop}%
\bibitem [{\citenamefont {Given}\ and\ \citenamefont
  {Stell}(1992)}]{GivSte92JCP}%
  \BibitemOpen
  \bibfield  {author} {\bibinfo {author} {\bibfnamefont {J.~A.}\ \bibnamefont
  {Given}}\ and\ \bibinfo {author} {\bibfnamefont {G.}~\bibnamefont {Stell}},\
  }\href {\doibase 10.1063/1.463883} {\bibfield  {journal} {\bibinfo  {journal}
  {J. Chem. Phys.}\ }\textbf {\bibinfo {volume} {97}},\ \bibinfo {pages} {4573}
  (\bibinfo {year} {1992})}\BibitemShut {NoStop}%
\bibitem [{\citenamefont {Lomba}\ \emph {et~al.}(1993)\citenamefont {Lomba},
  \citenamefont {Given}, \citenamefont {Stell}, \citenamefont {Weis},\ and\
  \citenamefont {Levesque}}]{LomGivSteWeiLev93PRE}%
  \BibitemOpen
  \bibfield  {author} {\bibinfo {author} {\bibfnamefont {E.}~\bibnamefont
  {Lomba}}, \bibinfo {author} {\bibfnamefont {J.~A.}\ \bibnamefont {Given}},
  \bibinfo {author} {\bibfnamefont {G.}~\bibnamefont {Stell}}, \bibinfo
  {author} {\bibfnamefont {J.~J.}\ \bibnamefont {Weis}}, \ and\ \bibinfo
  {author} {\bibfnamefont {D.}~\bibnamefont {Levesque}},\ }\href {\doibase
  10.1103/PhysRevE.48.233} {\bibfield  {journal} {\bibinfo  {journal} {Phys.
  Rev. E}\ }\textbf {\bibinfo {volume} {48}},\ \bibinfo {pages} {233} (\bibinfo
  {year} {1993})}\BibitemShut {NoStop}%
\bibitem [{\citenamefont {Given}\ and\ \citenamefont
  {Stell}(1994)}]{GivSte94PA}%
  \BibitemOpen
  \bibfield  {author} {\bibinfo {author} {\bibfnamefont {J.~A.}\ \bibnamefont
  {Given}}\ and\ \bibinfo {author} {\bibfnamefont {G.}~\bibnamefont {Stell}},\
  }\href {\doibase 10.1016/0378-4371(94)90200-3} {\bibfield  {journal}
  {\bibinfo  {journal} {Physica A}\ }\textbf {\bibinfo {volume} {209}},\
  \bibinfo {pages} {495} (\bibinfo {year} {1994})}\BibitemShut {NoStop}%
\bibitem [{\citenamefont {Rosinberg}, \citenamefont {Tarjus},\ and\
  \citenamefont {Stell}(1994)}]{RosTarSte94JCP}%
  \BibitemOpen
  \bibfield  {author} {\bibinfo {author} {\bibfnamefont {M.~L.}\ \bibnamefont
  {Rosinberg}}, \bibinfo {author} {\bibfnamefont {G.}~\bibnamefont {Tarjus}}, \
  and\ \bibinfo {author} {\bibfnamefont {G.}~\bibnamefont {Stell}},\ }\href
  {\doibase 10.1063/1.467182} {\bibfield  {journal} {\bibinfo  {journal} {J.
  Chem. Phys.}\ }\textbf {\bibinfo {volume} {100}},\ \bibinfo {pages} {5172}
  (\bibinfo {year} {1994})}\BibitemShut {NoStop}%
\bibitem [{\citenamefont {Dong}, \citenamefont {Kierlik},\ and\ \citenamefont
  {Rosinberg}(1994)}]{DonKieRos94PRE}%
  \BibitemOpen
  \bibfield  {author} {\bibinfo {author} {\bibfnamefont {W.}~\bibnamefont
  {Dong}}, \bibinfo {author} {\bibfnamefont {E.}~\bibnamefont {Kierlik}}, \
  and\ \bibinfo {author} {\bibfnamefont {M.~L.}\ \bibnamefont {Rosinberg}},\
  }\href@noop {} {\bibfield  {journal} {\bibinfo  {journal} {Phys. Rev. E}\
  }\textbf {\bibinfo {volume} {50}},\ \bibinfo {pages} {4750} (\bibinfo {year}
  {1994})}\BibitemShut {NoStop}%
\bibitem [{\citenamefont {Schmidt}(2002)}]{Sch02PRE}%
  \BibitemOpen
  \bibfield  {author} {\bibinfo {author} {\bibfnamefont {M.}~\bibnamefont
  {Schmidt}},\ }\href {\doibase 10.1103/PhysRevE.66.041108} {\bibfield
  {journal} {\bibinfo  {journal} {Phys. Rev. E}\ }\textbf {\bibinfo {volume}
  {66}},\ \bibinfo {pages} {041108} (\bibinfo {year} {2002})}\BibitemShut
  {NoStop}%
\bibitem [{\citenamefont {Schmidt}(2003)}]{Sch03PRE}%
  \BibitemOpen
  \bibfield  {author} {\bibinfo {author} {\bibfnamefont {M.}~\bibnamefont
  {Schmidt}},\ }\href@noop {} {\bibfield  {journal} {\bibinfo  {journal} {Phys.
  Rev. E}\ }\textbf {\bibinfo {volume} {68}},\ \bibinfo {pages} {021106}
  (\bibinfo {year} {2003})}\BibitemShut {NoStop}%
\bibitem [{\citenamefont {Reich}\ and\ \citenamefont
  {Schmidt}(2004)}]{ReiSch04JSP}%
  \BibitemOpen
  \bibfield  {author} {\bibinfo {author} {\bibfnamefont {H.}~\bibnamefont
  {Reich}}\ and\ \bibinfo {author} {\bibfnamefont {M.}~\bibnamefont
  {Schmidt}},\ }\href {\doibase 10.1023/B:JOSS.0000041752.55138.0a} {\bibfield
  {journal} {\bibinfo  {journal} {J. Stat. Phys.}\ }\textbf {\bibinfo {volume}
  {116}},\ \bibinfo {pages} {1683} (\bibinfo {year} {2004})}\BibitemShut
  {NoStop}%
\bibitem [{\citenamefont {Schmidt}(2009)}]{Sch09PRE}%
  \BibitemOpen
  \bibfield  {author} {\bibinfo {author} {\bibfnamefont {M.}~\bibnamefont
  {Schmidt}},\ }\href {\doibase 10.1103/PhysRevE.79.031405} {\bibfield
  {journal} {\bibinfo  {journal} {Phys. Rev. E}\ }\textbf {\bibinfo {volume}
  {79}},\ \bibinfo {pages} {031405} (\bibinfo {year} {2009})}\BibitemShut
  {NoStop}%
\bibitem [{\citenamefont {Lafuente}\ and\ \citenamefont
  {Cuesta}(2006)}]{LafCue06PRE}%
  \BibitemOpen
  \bibfield  {author} {\bibinfo {author} {\bibfnamefont {L.}~\bibnamefont
  {Lafuente}}\ and\ \bibinfo {author} {\bibfnamefont {J.~A.}\ \bibnamefont
  {Cuesta}},\ }\href {\doibase 10.1103/PhysRevE.74.041502} {\bibfield
  {journal} {\bibinfo  {journal} {Phys. Rev. E}\ }\textbf {\bibinfo {volume}
  {74}},\ \bibinfo {pages} {041502} (\bibinfo {year} {2006})}\BibitemShut
  {NoStop}%
\bibitem [{\citenamefont {Van~Tassel}\ \emph {et~al.}(1997)\citenamefont
  {Van~Tassel}, \citenamefont {Talbot}, \citenamefont {Viot},\ and\
  \citenamefont {Tarjus}}]{TasTalVioTar97PRE}%
  \BibitemOpen
  \bibfield  {author} {\bibinfo {author} {\bibfnamefont {P.~R.}\ \bibnamefont
  {Van~Tassel}}, \bibinfo {author} {\bibfnamefont {J.}~\bibnamefont {Talbot}},
  \bibinfo {author} {\bibfnamefont {P.}~\bibnamefont {Viot}}, \ and\ \bibinfo
  {author} {\bibfnamefont {G.}~\bibnamefont {Tarjus}},\ }\href {\doibase
  10.1103/PhysRevE.56.R1299} {\bibfield  {journal} {\bibinfo  {journal} {Phys.
  Rev. E}\ }\textbf {\bibinfo {volume} {56}},\ \bibinfo {pages} {R1299}
  (\bibinfo {year} {1997})}\BibitemShut {NoStop}%
\bibitem [{\citenamefont {Van~Tassel}(1997)}]{Tas97JCP}%
  \BibitemOpen
  \bibfield  {author} {\bibinfo {author} {\bibfnamefont {P.~R.}\ \bibnamefont
  {Van~Tassel}},\ }\href {\doibase 10.1063/1.475249} {\bibfield  {journal}
  {\bibinfo  {journal} {J. Chem. Phys.}\ }\textbf {\bibinfo {volume} {107}},\
  \bibinfo {pages} {9530} (\bibinfo {year} {1997})}\BibitemShut {NoStop}%
\bibitem [{\citenamefont {Van~Tassel}(1999)}]{Tas99PRE}%
  \BibitemOpen
  \bibfield  {author} {\bibinfo {author} {\bibfnamefont {P.~R.}\ \bibnamefont
  {Van~Tassel}},\ }\href {\doibase 10.1103/PhysRevE.60.R25} {\bibfield
  {journal} {\bibinfo  {journal} {Phys. Rev. E}\ }\textbf {\bibinfo {volume}
  {60}},\ \bibinfo {pages} {R25} (\bibinfo {year} {1999})}\BibitemShut
  {NoStop}%
\bibitem [{\citenamefont {Zhang}\ and\ \citenamefont
  {Van~Tassel}(2000{\natexlab{a}})}]{ZhaTas00JCP}%
  \BibitemOpen
  \bibfield  {author} {\bibinfo {author} {\bibfnamefont {L.}~\bibnamefont
  {Zhang}}\ and\ \bibinfo {author} {\bibfnamefont {P.~R.}\ \bibnamefont
  {Van~Tassel}},\ }\href {\doibase 10.1063/1.480874} {\bibfield  {journal}
  {\bibinfo  {journal} {J. Chem. Phys.}\ }\textbf {\bibinfo {volume} {112}},\
  \bibinfo {pages} {3006} (\bibinfo {year} {2000}{\natexlab{a}})}\BibitemShut
  {NoStop}%
\bibitem [{\citenamefont {Zhang}\ and\ \citenamefont
  {Van~Tassel}(2000{\natexlab{b}})}]{ZhaTas00MP}%
  \BibitemOpen
  \bibfield  {author} {\bibinfo {author} {\bibfnamefont {L.}~\bibnamefont
  {Zhang}}\ and\ \bibinfo {author} {\bibfnamefont {P.~R.}\ \bibnamefont
  {Van~Tassel}},\ }\href {\doibase 10.1080/00268970009483357} {\bibfield
  {journal} {\bibinfo  {journal} {Mol. Phys.}\ }\textbf {\bibinfo {volume}
  {98}},\ \bibinfo {pages} {1521} (\bibinfo {year}
  {2000}{\natexlab{b}})}\BibitemShut {NoStop}%
\bibitem [{\citenamefont {Zhang}, \citenamefont {Cheng},\ and\ \citenamefont
  {Van~Tassel}(2001)}]{ZhaCheTas01PRE}%
  \BibitemOpen
  \bibfield  {author} {\bibinfo {author} {\bibfnamefont {L.}~\bibnamefont
  {Zhang}}, \bibinfo {author} {\bibfnamefont {S.}~\bibnamefont {Cheng}}, \ and\
  \bibinfo {author} {\bibfnamefont {P.~R.}\ \bibnamefont {Van~Tassel}},\ }\href
  {\doibase 10.1103/PhysRevE.64.042101} {\bibfield  {journal} {\bibinfo
  {journal} {Phys. Rev. E}\ }\textbf {\bibinfo {volume} {64}},\ \bibinfo
  {pages} {042101} (\bibinfo {year} {2001})}\BibitemShut {NoStop}%
\bibitem [{\citenamefont {Hansen}\ and\ \citenamefont
  {McDonald}(2006)}]{macdohansen3ed}%
  \BibitemOpen
  \bibfield  {author} {\bibinfo {author} {\bibfnamefont {J.-P.}\ \bibnamefont
  {Hansen}}\ and\ \bibinfo {author} {\bibfnamefont {I.~R.}\ \bibnamefont
  {McDonald}},\ }\href@noop {} {\emph {\bibinfo {title} {Theory of simple
  liquids, Third edition}}}\ (\bibinfo  {publisher} {Academic Press},\ \bibinfo
  {address} {London},\ \bibinfo {year} {2006})\BibitemShut {NoStop}%
\bibitem [{Note1()}]{Note1}%
  \BibitemOpen
  \bibinfo {note} {These quantities are defined in terms of the usual one- and
  two-body density operators of the theory of mixtures, with configurational
  averages taken over the composite probability distribution $\protect \mathcal
  {P}_\protect \text {mt}(N_\protect \text {m}, \protect \mathbf
  {q}^{N_\protect \text {m}}, N_\protect \text {t}, \protect \mathbf
  {s}^{N_\protect \text {t}}) \protect \mathcal {P}_\protect \text
  {f}(N_\protect \text {f}, \protect \mathbf {r}^{N_\protect \text {f}} |
  N_\protect \text {m}, \protect \mathbf {q}^{N_\protect \text {m}})$. See
  Ref.~\protect \rev@citealpnum {Kra10PRE} for details.}\BibitemShut {Stop}%
\bibitem [{Note2()}]{Note2}%
  \BibitemOpen
  \bibinfo {note} {The introduction of these indicator functions is an
  alternative to the splitting of the configurational integrals used by
  Scheidler \protect \emph {et al.} in Ref.~\protect \rev@citealpnum
  {SchKobBin04JPCB}.}\BibitemShut {Stop}%
\bibitem [{\citenamefont {Meroni}, \citenamefont {Levesque},\ and\
  \citenamefont {Weis}(1996)}]{MerLevWei96JCP}%
  \BibitemOpen
  \bibfield  {author} {\bibinfo {author} {\bibfnamefont {A.}~\bibnamefont
  {Meroni}}, \bibinfo {author} {\bibfnamefont {D.}~\bibnamefont {Levesque}}, \
  and\ \bibinfo {author} {\bibfnamefont {J.~J.}\ \bibnamefont {Weis}},\ }\href
  {\doibase 10.1063/1.471954} {\bibfield  {journal} {\bibinfo  {journal} {J.
  Chem. Phys.}\ }\textbf {\bibinfo {volume} {105}},\ \bibinfo {pages} {1101}
  (\bibinfo {year} {1996})}\BibitemShut {NoStop}%
\bibitem [{\citenamefont {Schwanzer}\ \emph {et~al.}(2009)\citenamefont
  {Schwanzer}, \citenamefont {Coslovich}, \citenamefont {Kurzidim},\ and\
  \citenamefont {Kahl}}]{SchCosKurKah09MP}%
  \BibitemOpen
  \bibfield  {author} {\bibinfo {author} {\bibfnamefont {D.~F.}\ \bibnamefont
  {Schwanzer}}, \bibinfo {author} {\bibfnamefont {D.}~\bibnamefont
  {Coslovich}}, \bibinfo {author} {\bibfnamefont {J.}~\bibnamefont {Kurzidim}},
  \ and\ \bibinfo {author} {\bibfnamefont {G.}~\bibnamefont {Kahl}},\ }\href
  {\doibase 10.1080/00268970902845321} {\bibfield  {journal} {\bibinfo
  {journal} {Mol. Phys.}\ }\textbf {\bibinfo {volume} {107}},\ \bibinfo {pages}
  {433} (\bibinfo {year} {2009})}\BibitemShut {NoStop}%
\bibitem [{\citenamefont {Karmakar}\ and\ \citenamefont
  {Parisi}(2013)}]{KarPar13PNAS}%
  \BibitemOpen
  \bibfield  {author} {\bibinfo {author} {\bibfnamefont {S.}~\bibnamefont
  {Karmakar}}\ and\ \bibinfo {author} {\bibfnamefont {G.}~\bibnamefont
  {Parisi}},\ }\href@noop {} {\bibfield  {journal} {\bibinfo  {journal} {Proc.
  Natl. Acad. Sci. U.S.A.}\ }\textbf {\bibinfo {volume} {110}},\ \bibinfo
  {pages} {2752} (\bibinfo {year} {2013})}\BibitemShut {NoStop}%
\bibitem [{Note3()}]{Note3}%
  \BibitemOpen
  \bibinfo {note} {Possibilities also exist in the framework of the density
  functional theory, \cite {Sch02PRE,Sch03PRE,ReiSch04JSP,Sch09PRE,LafCue06PRE}
  through the use of specially tailored test-particle methods, \cite {Sch09PRE}
  but they have been little explored.}\BibitemShut {Stop}%
\bibitem [{Note4()}]{Note4}%
  \BibitemOpen
  \bibinfo {note} {There is no infinite potential excluding the fluid from the
  matrix in the original work reported in Ref.~\protect \rev@citealpnum
  {DonKieRos94PRE}, but one can easily add it to the formalism and check that
  this does not change the equations. For definiteness, it should also be
  mentioned that the species referred to as the template in the present work
  corresponds in Ref.~\protect \rev@citealpnum {DonKieRos94PRE} to the
  ``ghost'' matrix particles that do not interact with the fluid.}\BibitemShut
  {Stop}%
\bibitem [{\citenamefont {Attard}(2002)}]{Att02book}%
  \BibitemOpen
  \bibfield  {author} {\bibinfo {author} {\bibfnamefont {P.}~\bibnamefont
  {Attard}},\ }\href@noop {} {\emph {\bibinfo {title} {Thermodynamics and
  Statistical Mechanics - Equilibrium by Entropy Maximisation}}}\ (\bibinfo
  {publisher} {Academic Press},\ \bibinfo {address} {London},\ \bibinfo {year}
  {2002})\BibitemShut {NoStop}%
\bibitem [{\citenamefont {Madden}(1995)}]{Mad95JCPcom}%
  \BibitemOpen
  \bibfield  {author} {\bibinfo {author} {\bibfnamefont {W.~G.}\ \bibnamefont
  {Madden}},\ }\href {\doibase 10.1063/1.469287} {\bibfield  {journal}
  {\bibinfo  {journal} {J. Chem. Phys.}\ }\textbf {\bibinfo {volume} {102}},\
  \bibinfo {pages} {5572} (\bibinfo {year} {1995})}\BibitemShut {NoStop}%
\bibitem [{Note5()}]{Note5}%
  \BibitemOpen
  \bibinfo {note} {The key aspect is that, in the diagrams contributing to the
  disconnected correlation functions, all paths linking the fluid root points
  must go through at least one matrix field point. Since nodal points are
  allowed in $h_\protect \text {dis}$, it is possible to have only one matrix
  field point, which is precisely and necessarily nodal. On the other hand,
  nodal points being forbidden in $c_\protect \text {dis}$, at least two matrix
  field points must be present.}\BibitemShut {Stop}%
\bibitem [{Note6()}]{Note6}%
  \BibitemOpen
  \bibinfo {note} {In the more recent Refs.~\protect \rev@citealpnum
  {ChaChaTar13JCP} and \protect \rev@citealpnum {ChaTar13PRE}, the full length
  $(x \rho )^{-1/3}$ is used as the variable in similar plots. We do not do so
  here, because the resulting horizontal shift with $\rho $ of the endpoints
  corresponding to $x=1$ leads to overlapping curves and a less legible
  figure.}\BibitemShut {Stop}%
\bibitem [{\citenamefont {Cardenas}, \citenamefont {Franz},\ and\ \citenamefont
  {Parisi}(1999)}]{CarFraPar99JCP}%
  \BibitemOpen
  \bibfield  {author} {\bibinfo {author} {\bibfnamefont {M.}~\bibnamefont
  {Cardenas}}, \bibinfo {author} {\bibfnamefont {S.}~\bibnamefont {Franz}}, \
  and\ \bibinfo {author} {\bibfnamefont {G.}~\bibnamefont {Parisi}},\
  }\href@noop {} {\bibfield  {journal} {\bibinfo  {journal} {J. Chem. Phys.}\
  }\textbf {\bibinfo {volume} {110}},\ \bibinfo {pages} {1726} (\bibinfo {year}
  {1999})}\BibitemShut {NoStop}%
\bibitem [{\citenamefont {Parisi}\ and\ \citenamefont
  {Zamponi}(2010)}]{ParZam10RMP}%
  \BibitemOpen
  \bibfield  {author} {\bibinfo {author} {\bibfnamefont {G.}~\bibnamefont
  {Parisi}}\ and\ \bibinfo {author} {\bibfnamefont {F.}~\bibnamefont
  {Zamponi}},\ }\href@noop {} {\bibfield  {journal} {\bibinfo  {journal} {Rev.
  Mod. Phys.}\ }\textbf {\bibinfo {volume} {82}},\ \bibinfo {pages} {789}
  (\bibinfo {year} {2010})}\BibitemShut {NoStop}%
\bibitem [{\citenamefont {Lavis}\ and\ \citenamefont
  {Bell}(1999)}]{LavBel99bookv1}%
  \BibitemOpen
  \bibfield  {author} {\bibinfo {author} {\bibfnamefont {D.~A.}\ \bibnamefont
  {Lavis}}\ and\ \bibinfo {author} {\bibfnamefont {G.~M.}\ \bibnamefont
  {Bell}},\ }\href@noop {} {\emph {\bibinfo {title} {Statistical Mechanics of
  Lattice Systems. Volume 1: Closed-Form and Exact Solutions}}}\ (\bibinfo
  {publisher} {Springer-Verlag},\ \bibinfo {address} {Berlin},\ \bibinfo {year}
  {1999})\BibitemShut {NoStop}%
\bibitem [{Note7()}]{Note7}%
  \BibitemOpen
  \bibinfo {note} {Another possibility is to introduce matrix occupancy
  variables, $(u_1,u_2,\protect \ldots ,u_V)$, such that $u_i=1$ if $i\in
  \protect \{i_\protect \text {m}\protect \}$ and $u_i=0$ otherwise. One then
  obtains the Hamiltonian \begin {multline} H_\protect \text {f}(\protect
  \mathbf {n}^{V}|\protect \mathbf {u}^{V}) = \\ - w \DOTSB \sum@ \slimits@
  _{\delimiter "426830A i j \delimiter "526930B } n_i n_j (1-u_i) (1-u_j) + n_i
  (1-u_i) u_j + n_j (1-u_j) u_i \\ - \left [ \mu + k_\protect \text {B}T
  \protect \qopname \relax o{ln}(1-x) \right ] \DOTSB \sum@ \slimits@ _{i} n_i
  (1-u_i). \end {multline} This formulation shows that the present PP system
  bears strong connection to a lattice-gas model previously studied by
  Rosinberg \protect \emph {et al.}, who showed that it can be readily
  transformed into an Ising model with correlated site dilution and random
  fields. \cite {PitRosSteTar95PRL,PitRosTar96MS,KieRosTar97JSP,
  KieRosTarPit98MP}}\BibitemShut {NoStop}%
\bibitem [{\citenamefont {Grinstein}\ and\ \citenamefont
  {Mukamel}(1983)}]{GriMuk83PRB}%
  \BibitemOpen
  \bibfield  {author} {\bibinfo {author} {\bibfnamefont {G.}~\bibnamefont
  {Grinstein}}\ and\ \bibinfo {author} {\bibfnamefont {D.}~\bibnamefont
  {Mukamel}},\ }\href@noop {} {\bibfield  {journal} {\bibinfo  {journal} {Phys.
  Rev. B}\ }\textbf {\bibinfo {volume} {27}},\ \bibinfo {pages} {4503}
  (\bibinfo {year} {1983})}\BibitemShut {NoStop}%
\bibitem [{\citenamefont {Lee}\ and\ \citenamefont {Yang}(1952)}]{LeeYan52PR}%
  \BibitemOpen
  \bibfield  {author} {\bibinfo {author} {\bibfnamefont {T.~D.}\ \bibnamefont
  {Lee}}\ and\ \bibinfo {author} {\bibfnamefont {C.~N.}\ \bibnamefont {Yang}},\
  }\href {\doibase 10.1103/PhysRev.87.410} {\bibfield  {journal} {\bibinfo
  {journal} {Phys. Rev.}\ }\textbf {\bibinfo {volume} {87}},\ \bibinfo {pages}
  {410} (\bibinfo {year} {1952})}\BibitemShut {NoStop}%
\bibitem [{Note8()}]{Note8}%
  \BibitemOpen
  \bibinfo {note} {Incidentally, this result implies that a homogeneous PP
  system is never in equilibrium with the bulk fluid from which it has been
  prepared. If the two are brought into contact, particles will spontaneously
  move from the bulk to the porous matrix, until equalization of the chemical
  potentials is achieved.}\BibitemShut {Stop}%
\bibitem [{\citenamefont {Pitard}\ \emph {et~al.}(1995)\citenamefont {Pitard},
  \citenamefont {Rosinberg}, \citenamefont {Stell},\ and\ \citenamefont
  {Tarjus}}]{PitRosSteTar95PRL}%
  \BibitemOpen
  \bibfield  {author} {\bibinfo {author} {\bibfnamefont {E.}~\bibnamefont
  {Pitard}}, \bibinfo {author} {\bibfnamefont {M.~L.}\ \bibnamefont
  {Rosinberg}}, \bibinfo {author} {\bibfnamefont {G.}~\bibnamefont {Stell}}, \
  and\ \bibinfo {author} {\bibfnamefont {G.}~\bibnamefont {Tarjus}},\ }\href
  {\doibase 10.1103/PhysRevLett.74.4361} {\bibfield  {journal} {\bibinfo
  {journal} {Phys. Rev. Lett.}\ }\textbf {\bibinfo {volume} {74}},\ \bibinfo
  {pages} {4361} (\bibinfo {year} {1995})}\BibitemShut {NoStop}%
\bibitem [{\citenamefont {Pitard}, \citenamefont {Rosinberg},\ and\
  \citenamefont {Tarjus}(1996)}]{PitRosTar96MS}%
  \BibitemOpen
  \bibfield  {author} {\bibinfo {author} {\bibfnamefont {E.}~\bibnamefont
  {Pitard}}, \bibinfo {author} {\bibfnamefont {M.~L.}\ \bibnamefont
  {Rosinberg}}, \ and\ \bibinfo {author} {\bibfnamefont {G.}~\bibnamefont
  {Tarjus}},\ }\href@noop {} {\bibfield  {journal} {\bibinfo  {journal} {Mol.
  Simul.}\ }\textbf {\bibinfo {volume} {17}},\ \bibinfo {pages} {399} (\bibinfo
  {year} {1996})}\BibitemShut {NoStop}%
\bibitem [{\citenamefont {Kierlik}, \citenamefont {Rosinberg},\ and\
  \citenamefont {Tarjus}(1997)}]{KieRosTar97JSP}%
  \BibitemOpen
  \bibfield  {author} {\bibinfo {author} {\bibfnamefont {E.}~\bibnamefont
  {Kierlik}}, \bibinfo {author} {\bibfnamefont {M.~L.}\ \bibnamefont
  {Rosinberg}}, \ and\ \bibinfo {author} {\bibfnamefont {G.}~\bibnamefont
  {Tarjus}},\ }\href {\doibase 10.1007/BF02770762} {\bibfield  {journal}
  {\bibinfo  {journal} {J. Stat. Phys.}\ }\textbf {\bibinfo {volume} {89}},\
  \bibinfo {pages} {215} (\bibinfo {year} {1997})}\BibitemShut {NoStop}%
\bibitem [{\citenamefont {Kierlik}\ \emph {et~al.}(1998)\citenamefont
  {Kierlik}, \citenamefont {Rosinberg}, \citenamefont {Tarjus},\ and\
  \citenamefont {Pitard}}]{KieRosTarPit98MP}%
  \BibitemOpen
  \bibfield  {author} {\bibinfo {author} {\bibfnamefont {E.}~\bibnamefont
  {Kierlik}}, \bibinfo {author} {\bibfnamefont {M.~L.}\ \bibnamefont
  {Rosinberg}}, \bibinfo {author} {\bibfnamefont {G.}~\bibnamefont {Tarjus}}, \
  and\ \bibinfo {author} {\bibfnamefont {E.}~\bibnamefont {Pitard}},\
  }\href@noop {} {\bibfield  {journal} {\bibinfo  {journal} {Mol. Phys.}\
  }\textbf {\bibinfo {volume} {95}},\ \bibinfo {pages} {341} (\bibinfo {year}
  {1998})}\BibitemShut {NoStop}%
\end{thebibliography}

%

\end{document}